\documentclass[final,leqno,onefignum,onetabnum,10pt,a4paper,twocolumn]{article}
\usepackage[a4paper, top=1in, bottom=1.25in, left=0.7in, right=0.7in]{geometry}
\usepackage{graphicx}
\usepackage{hyperref}
\usepackage{mathptmx}      
\usepackage[misc,geometry]{ifsym}	
\usepackage[super]{nth}				
\usepackage{amsmath}			
\usepackage{amssymb}
\usepackage[numbers,sort&compress]{natbib}	
\usepackage{booktabs}
\usepackage{subfig}		
\usepackage[heightadjust=all,valign=b]{floatrow}
\usepackage{fr-subfig}
\usepackage{caption}
\usepackage{tikz}
\usepackage{pgfplots}
\pgfplotsset{compat=1.9}
\usepackage{tikzexternal}
\pgfplotstableread{
0             0.00000111  0.00000445  0.00011092   0.0004335   0.0110145     0.04296      0.1533 
0.1           0.00000111  0.00000443  0.00010837   0.0004224   0.0107571     0.04232      0.1518 
0.25          0.00000110  0.00000441  0.00010574   0.0004207   0.0104161     0.04142      0.1497 
0.5           0.00000110  0.00000437  0.00010423   0.0004166   0.0101738     0.04000      0.1465 
1             0.00000108  0.00000430  0.00010197   0.0004084   0.0098010     0.03797      0.1405 
2.5           0.00000104  0.00000414  0.00009742   0.0003900   0.0092417     0.03457      0.1262 
5             0.00000098  0.00000393  0.00009217   0.0003698   0.0086633     0.03189      0.1127 
10            0.00000089  0.00000363  0.00008627   0.0003444   0.0079854     0.02908      0.1003 
25            nan         nan         0.00009052   0.0003106   0.0073465     0.02662      nan          
50            nan         nan         nan	   nan         0.0071583     0.02633      0.0895 
100           nan         nan         nan          nan         0.0066882     0.02510      0.0877 
}\dataDiffMine
\pgfplotstableread{
0.1           0.01        nan         nan         nan         nan
0.5           0.01        0.10        nan         0.40        nan
1             nan         nan         0.16        nan         0.67
2             0.01        nan         0.16        nan         nan
5             0.01        0.10        0.16        nan         nan
10            nan         nan         0.16        nan         nan
2             nan         0.10        0.16        nan         nan
50            0.01        0.10        0.16        0.40        nan
100           nan         nan         0.16        nan         0.67
}\dataDiffSzSn
\pgfplotstableread{
0.0005		0.00000108	nan
0.001		0.00000430	nan
0.005		0.00010197	0.01
0.01		0.0004084	nan
0.05		0.0098010	0.10
0.1		0.03797		0.16
0.2		0.1405		nan
0.5		nan		0.40
1		nan		0.67
}\dataDiffEpsCompare
\pgfplotstableread{
0	0.3575	0.3575	0.3575	0.3575	0.3575
0.3	0.3837	0.3837	0.3837	0.3837	0.3837
0.6	0.4165	0.4165	0.4165	0.4165	0.4165
0.8	0.4438	0.4438	0.4438	0.4438	0.4438
1	0.4776	0.4776	0.4776	0.4776	0.4776
1.2	0.5209	0.5214	0.5214	0.5214	0.5214
1.4	0.583	0.5828	0.5828	0.5828	0.5828
1.6	0.6854	0.6854	0.6854	0.6854	0.6854
1.7	0.7837	0.7837	0.7837	0.7837	0.7687
1.71	0.7987	nan     nan     0.7987	0.7987
1.72	0.8159	nan     nan    	0.8159	0.8159
1.73	0.8362	nan     nan    	0.8363	0.8363
1.74	0.8633	nan     nan    	0.8633	0.8633
1.741	0.8668	nan     nan    0.8386   0.8386   	
1.742	0.8707	nan     nan    0.8410   0.8410 	
1.743	0.8762	nan     nan    0.8434   0.8756    	
1.74301	0.8761	nan     nan    nan     nan    	
1.74302	0.8764	nan     nan    nan     nan    	
1.74303	0.8764	nan     nan    nan     nan    	
1.74304	0.8763	nan     nan    nan     nan    	
1.74305	0.8765	nan     nan    nan     nan    	
1.74306	0.8766	nan     nan    nan     nan    	
1.74307	0.8767	nan     nan    nan     nan    	
1.74309	0.8768	nan     nan    nan     nan    	
1.7431	0.8762	nan     nan    nan     nan    	
1.74311	0.8768	nan     nan    nan     nan    	
1.74312	0.8769	nan     nan    nan     nan    	
1.74313	0.8765	nan     nan    nan     nan    	
1.74314	0.877	nan     nan    nan     nan    	
1.74315	0.8767	nan     nan    	nan     nan    
1.75	nan	0.8857	0.9056	nan     nan    
1.76	nan	0.9142	0.9529	nan     nan    
1.77	nan	0.9449	1.004	nan     nan    
1.78	nan	0.9773	1.064	nan     nan    
1.79	nan	1.012	1.129	nan     nan    
1.8	nan	1.05	1.196	nan     nan    
1.81	nan	nan	1.275	nan     nan    
1.82	nan	nan	1.356	nan     nan    
1.83	nan	nan	1.441	nan     nan    
1.84	nan	nan	1.541	nan     nan    
1.85	nan	1.274	1.648	nan     nan    
1.87	nan	nan	1.888	nan     nan    
1.9	nan	1.563	nan	nan     nan    
2	nan	2.385	nan	nan     nan    
2.1	nan	3.584	nan	nan     nan    
}\dataCutoffFy
\pgfplotstableread{
0	-1.159	-1.159	-1.159	-1.159	-1.159	-1.159
0.3	-1.301	-1.301	-1.301	-1.301	-1.301	-1.301
0.6	-1.496	-1.496	-1.496	-1.496	-1.496	-1.496
0.8	-1.675	-1.675	-1.675	-1.675	-1.675	-1.675
1	-1.922	-1.922	-1.922	-1.922	-1.922	-1.922
1.2	-2.288	-2.291	-2.291	-2.291	-2.291	-2.291
1.4	-2.93	nan	-2.927	-2.927	-2.927	-2.927
1.6	-4.485	nan	-4.485	-4.485	-4.485	-4.485
1.7	-7.151	nan	-7.151	-7.151	-7.151	-7.151
1.71	-7.754	nan	-7.754	-7.754	-7.754	-7.754
1.72	-8.549	nan	-8.549	-8.549	-8.549	-8.549
1.73	-9.686	nan	-9.686	-9.686	-9.686	-9.686
1.731	nan	nan	-9.784	nan	nan	nan
1.733	nan	nan	-9.998	nan	nan	nan
1.74	-11.7	-9.535	-10.77	-9.831	nan	-11.7
1.741	-12.04	nan	nan	nan	-9.831	-9.831
1.742	-12.45	nan	-10.96	-9.984	-9.984	-9.984
1.743	-13.1	nan	-11.05	-10.15	-10.015	nan
1.7431	-13.2	nan	nan	nan	nan	nan
1.75	nan	nan	-11.67	-15.35	nan	nan
1.76	nan	nan	-12.5	-17.79	nan	nan
1.77	nan	nan	-13.28	-19.95	nan	nan
1.78	nan	-11.51	-14.09	-21.97	nan	nan
1.79	nan	-12	-14.87	-23.92	nan	nan
1.8	nan	nan	-15.65	-25.84	nan	nan
1.81	nan	nan	nan	-27.75	nan	nan
1.82	nan	nan	nan	-29.67	nan	nan
1.83	nan	nan	nan	-31.6	nan	nan
1.84	nan	nan	nan	-33.56	nan	nan
1.85	nan	nan	-19.66	-35.5	nan	nan
1.9	nan	nan	-23.98	nan	nan	nan
2	nan	nan	-34.14	nan	nan	nan
}\dataCutoffFx
\pgfplotstableread{
0	0.5625	0.5625	0.5625	0.5625	0.5625	0.5625
0.3	0.6161	0.6161	0.6161	0.6161	0.6161	0.6161
0.6	0.6863	0.6863	0.6863	0.6863	0.6863	0.6863
0.8	0.7474	0.7474	0.7474	0.7474	0.7474	0.7474
1	0.8269	0.8269	0.8269	0.8269	0.8269	0.8269
1.2	0.9347	0.9370	0.9370	0.9370	0.9370	0.9370
1.4	1.1074	nan	1.1058	1.1058	1.1058	1.1058
1.6	1.4446	nan	1.4446	1.4446	1.4446	1.4446
1.7	1.8723	nan	1.8723	1.8723	1.8723	1.8723
1.71	1.9533	nan	1.9532	1.9532	1.9532	1.9532
1.72	2.0541	nan	2.0541	2.0541	2.0541	2.0541
1.73	2.1904	nan	2.1904	2.1904	2.1904	2.1904
1.731	nan	nan	2.2001	nan	nan	nan
1.733	nan	nan	2.2259	nan	nan	nan
1.74	2.4271	2.2372	2.3312	6.8415	2.4269	2.4269
1.741	2.4704	nan	nan	nan	4.4146	4.4146
1.742	2.5284	nan	2.3591	4.4503	4.4503	6.9771
1.743	5.2922	nan	2.3741	4.4879	4.4879	2.2439
1.7431	5.3967	nan	nan	nan	nan	nan
1.75	nan	nan	2.4755	2.8193	nan	nan
1.76	nan	nan	2.6246	3.1324	nan	nan
1.77	nan	nan	2.7732	3.4640	nan	nan
1.78	nan	2.6573	2.9400	3.8090	nan	nan
1.79	nan	2.7706	3.1046	4.1686	nan	nan
1.8	nan	nan	3.2765	4.5325	nan	nan
1.81	nan	nan	nan	4.9274	nan	nan
1.82	nan	nan	nan	5.3377	nan	nan
1.83	nan	nan	nan	5.7535	nan	nan
1.84	nan	nan	nan	6.1966	nan	nan
1.85	nan	nan	4.2252	13.3380	nan	nan
1.9	nan	nan	5.3413	nan	nan	nan
2	nan	nan	8.1949	nan	nan	nan
}\dataCutoffPmax
\newcommand{\tens}{\pmb}
\newcommand{\vect}{\pmb}
\DeclareMathOperator{\divv}{div}	
\DeclareMathOperator{\tr}{tr}		
\newcommand{\ud}{\mathrm{d}}

\newcommand{\defeq}{\stackrel{\mathrm{def}}{=}}
\newcommand{\eps}{\varepsilon}
\newcommand{\Rey}{\mathrm{Re}}
\newcommand{\WOd}[3]{\vect{W}^{#1,#2}_{#3}(\Omega)}
\newcommand{\Section}[1]{Sect.~\ref{sec:#1}}
\newcommand{\Figure}[1]{Fig.~\ref{fig:#1}}
\newcommand{\Figurep}[1]{Fig.~\ref{plot:#1}}
\begin{document}

\title{%
Numerical simulations of an incompressible piezoviscous fluid 
flowing in a~plane slider bearing%
\thanks{%
Submitted to Meccanica on
January 11, 2017.
The final publication is available at Springer via 
\href{http://dx.doi.org/10.1007/s11012-017-0731-0}{http://dx.doi.org/10.1007/s11012-017-0731-0}.%
}%
}

\author{%
Martin Lanzend{\"o}rfer
\and
Josef M{\'a}lek
\and
Kumbakonam R.\ Rajagopal
\thanks{%
M.\ Lanzend{\"o}rfer, J.\ M{\'a}lek, 
	Mathematical Institute, Faculty of Mathematics and Physics, Charles University, 
 	Prague, Czech Republic.
J.~M{\'a}lek acknowledges the support of the ERC-CZ project LL1202 financed by M{\v{S}}MT (Ministry of Education, Youth and Sports of the Czech Republic).
K.~R.~Rajagopal,
 	Department of Mechanical Engineering, Texas A\&M University, United States.
K.~R.~Rajagopal thanks the National Science Foundation, United States for its support.
}%
}


\date{}

\maketitle

\begin{abstract}\small
We provide numerical simulations 
of an~incompressible pressure-thickening and shear-thinning lubricant
flowing in a~plane slider bearing.
We study the influence of several parameters, 
namely the ratio of the characteristic lengths $\eps>0$
(with $\eps\searrow0$ representing the Reynolds lubrication approximation);
the coefficient of the exponential pressure--viscosity relation $\alpha^*\geq0$;
the parameter $G^*\geq0$ related to the Carreau--Yasuda shear-thinning model
and the modified Reynolds number $\Rey_\eps\geq0$.
The finite element approximations to the steady isothermal flows are computed
without resorting to the lubrication approximation.
We obtain the numerical solutions as long as the variation of the viscous stress
$\tens{S}=2\eta(p,\tr\tens{D}^2)\tens{D}$ with the pressure is limited, say $|\partial\tens{S}/\partial p|\leq1$.
We show conclusively that the existing practice 
of avoiding the numerical difficulties by cutting the viscosity off for large pressures
leads to results that depend sorely on the artificial cut-off parameter.
We observe that the piezoviscous rheology generates pressure differences across the fluid film.
\end{abstract}

\section{Introduction} 
Lubrication problems represent a~set of important engineering applications 
that have been a~source of inspiration for a~great deal of research in fluid dynamics.
The plane slider flow described in the next section embodies 
a~classical prototype of 
hydrodynamic lubrication.
Two solid surfaces in relative motion are separated by a~thin layer of a~liquid lubricant,
the fluid film being thick enough still to separate the surfaces completely.
Since the fundamental work by Reynolds \cite{Reynolds_1886}, the \emph{lubrication approximation}
approach, which considerably reduces the system of equations governing the thin film flow,
proved to be a~very useful and flexible tool.

Within the class of lubrication problems, one that presents challenging issues is Elastohydrodynamic
Lubrication (EHL) wherein one encounters extremely high peak pressures%
\footnote{
The terminology ``pressure'' has been used to define a~variety of disparate quantities and can be a~source 
for a~great deal of confusion, especially when discussing lubricants since many lubricants that are used are 
non-Newtonian fluids (see \cite{Rajagopal2015} for a~detailed discussion of the concept of ``pressure''). 
In this study, ``pressure'' signifies the mean normal stress.
} of the order of a~{GPa}, very high shear rates, significant variations in temperature 
(see \cite{Bair2006,Bair_lubrication_book2007}), and deformation of the solid boundary 
(see \cite{Szeri_Lubrication_book1998}).
The competing effects of the increase of the viscosity due to the high pressures,
and the decrease of the viscosity due to the shear thinning at high shear rates as well as increases in
temperature present further challenges both with regard to rigorous mathematical and numerical
analysis, and computation.

Alongside the Reynolds approximation approach, which has served as the exclusive tool for engineering
predictions, the more general tools of Computational Fluid Dynamics (CFD) were brought to 
bear on lubrication problems recently, e.g.~\cite{
Almqvist2004,%
Almqvist2008,%
Almqvist2002,%
Bruyere2012,%
Hartinger2008,%
Knauf2013}.
CFD simulations are expected to allow one to get a~more detailed and accurate
understanding of the flow involved in the lubrication problems, that cannot be achieved within the context
of the Reynolds approximation, especially with regard to 
problems involving starved lubrication, 
problems involving rough and dimpled surfaces, cavitation,
or more complex rheology or geometry,
see~\cite{Lugt2011} for relevant references.
It is worth noting that the available numerical results based on 
solving the full system of equations governing the flow 
have not considered the heavily loaded regimes so far.

The present paper focuses on a~particular issue pointed out already 
by Bair et~al.\ \cite{BairKhonsariWiner_1998},
that 
the pressure-thickening response itself eventually causes the violation of the lubrication assumptions.
Namely, that a~gradient of pressure in the direction across the film is generated in the flow. 
The same observation has lead to the revision of Reynolds equation in the piezoviscous regime,
see~\cite{RajSzeri_2003,BayadaEtal_2013,GustafssonRajEtal_2015}.
Yet another important consequence of this rather specific feature of piezoviscous fluids
is that the momentum equation describing the flow exhibits structural changes,
once certain threshold of the pressure and shear rate is reached.
This is well reflected in the results that are available concerning 
the existence and uniqueness of weak solutions, which are based on assumptions that 
allow for the realistic pressure- and shear rate- viscosity relations 
only up to that threshold, see the references in \Section{numerics}.

Bearing this in mind, in contrast to the previous studies based on the CFD approach referred above,
we restrict ourselves to a~simpler setting.
This allows us to focus on some important issues, which we believe are characteristic of more realistic models as well 
but which have not been studied in the detail that they deserve in the literature so far.

In \Section{governing}
  we recall the partial differential equations governing the planar steady isothermal flow of a~homogeneous incompressible viscous fluid
  and 
  we develop the dimensionless governing equations within the context of the plane slider geometry.
  The boundary conditions for the inflow and outflow boundaries of the domain
  are discussed in \Section{bc} in detail.
  We describe the pressure-thickening and shear-thinning rheology provided by the Carreau--Yasuda relation
  with the exponential pressure--viscosity law.
  Such rheology is simple enough for the purpose of discussing how the dimensionless parameters affect the flow.
  At the same time, it provides a~realistic model that is not altered to fit into any class of constitutive relations
  accessible by the theoretical existence and uniqueness results available.
  
In \Section{numerics} 
  we introduce the finite element approximations to be used for carrying out the numerical simulations.
  We recall the current limitations of both the theoretical well-posedness results available 
  and of the numerical approach based on the Galerkin (finite element) approximations.
  We discuss the constraint with regard to the variation of the viscous stress with the pressure,
  which is observed in numerical experiments and is analogous to the assumptions needed 
  to establish existence results in the theoretical works.
  We are able to carry out the numerical simulations only within a~certain range of pressure and shear rate
  where the constraint is met.

 
\Section{results} starts by demonstrating the basic features of the flow in the case of Navier--Stokes fluid.
  Then we incorporate the pressure--viscosity relation into the problem and carry out a~set of numerical simulations 
  with $\eta=\eta_0 e^{\alpha p}$, $\alpha>0$.
  It is customary in numerical computations to avoid numerical difficulties 
  by cutting the viscosity off above given threshold for the pressure.
  We document by numerical calculations in \Section{cutoff},
  that such a~procedure may actually
  lead to very different results for the problem under consideration, depending on the cut-off parameter.
  Therefore, there is no cut-off utilised in the subsequent results presented in~\Section{results}.
  We show that the response of a~piezoviscous fluid leads to variations of pressure 
  across the film in \Section{dp}.
  Finally, we study the consequences of the fluid being shear-thinning
  and we also determine the effect of inertia on the characteristics of the flow.

\section{Setting of the mathematical problem}
\label{sec:governing}
\subsection{Governing equations}
\label{sec::governing}

We consider a~planar steady isothermal flow of a~homogeneous incompressible viscous fluid,
governed by the system of equations
\begin{equation}\label{eq:steady-incompressible}
\left.\begin{array}{rcl}
	\divv \vect{v} &=& 0
\\	\divv ( \rho \vect{v} \otimes \vect{v} ) - \divv \tens{T} &=& \vect{f}
\end{array}\right\}
\quad	\textrm{in }\Omega \subset \mathbb{R}^2,
\end{equation}
where the Cauchy stress tensor $\tens{T}$ is given 
by the relation
\begin{equation}\label{eq:T}
	\tens{T} = - p\,\tens{I} + 2 \eta ( p, \tr \tens{D}^2 )\tens{D}
.\end{equation}
The unknowns are the velocity $\vect{v}$ and the pressure $p$, 
while the given data are the density~$\rho$, the body force $\vect{f}$
and the relation $\eta=\eta(p,\tr\tens{D}^2)$, which characterizes the viscosity 
of a~pressure-thickening and shear-thinning lubricant. 
In the above equation $\tens{D} = \tfrac12( \nabla\vect{v} + (\nabla\vect{v})^T )$ 
denotes the symmetric part of the velocity gradient.
Note that here $p$ coincides with the mean value stress $m=-\tfrac12\tr\tens{T}$, by virtue of~$\tr\tens{D}=\divv\vect{v}=0$,
cf.~\cite{PrusaRaj_2013}.

The assumption of the flow being isothermal is made for the sake of simplicity.
Similarly, we do not allow for elastic deformation of the solid surfaces, 
so that a~flow in the fixed domain $\Omega$ is considered instead.
Note also that we implicitly assume that the resulting pressure field would remain positive throughout the domain, 
so that we need not discuss the possibility of cavitation within the flow.
Since we study the flow in between converging surfaces, the latter assumption is reasonable.

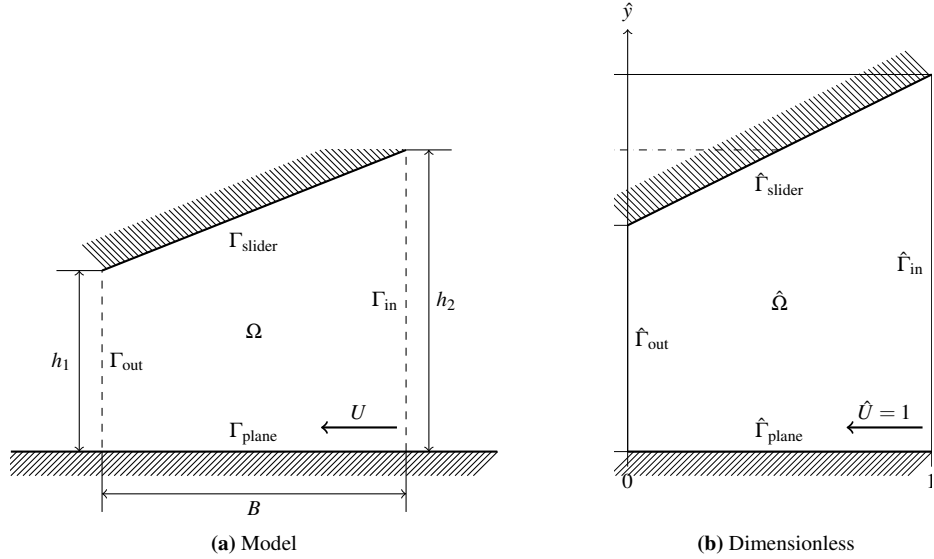
\begin{figure*}[t!]
\centering
\subfloat[Model]{%
\label{fig::geometry}
        \centering
	\begin{tikzpicture}[scale=0.4]
	\footnotesize
\coordinate (A) at (0,0) {};
\coordinate (B) at (10cm,0) {};
\coordinate (C) at (0,6cm) {};
\coordinate (D) at (10cm,10cm);
\node at (5cm,4cm){$\Omega$};

\path[coordinate] (A) -- (B) coordinate[pos=-0.3](Am) coordinate[pos=1.3](Bp);
\draw [thick] (Am) -- (Bp);
\draw [thick] (C) -- (D);
\draw [thin,dashed] (A) -- (B) node[pos=0.5,above]{$\Gamma_{\mathrm{plane}}$} %
		-- (D) node[pos=0.5,left]{$\Gamma_{\mathrm{in}}$} %
		-- (C) node[pos=0.5,below=0.2cm]{$\Gamma_{\mathrm{slider}}$} %
		-- (A) node[pos=0.5,right]{$\Gamma_{\mathrm{out}}$};
\coordinate (Cm) at (C);
\coordinate (Dm) at (D);
\foreach \x in {1,...,80}{
	\path[coordinate] (Am) -- ++(0.2cm,0) coordinate(Am);
	\draw [thin] (Am) -- ++(-0.8cm,-0.8cm);
}
\foreach \x in {1,...,66}{
	\path[coordinate] (Cm) -- (Dm) coordinate[pos=0.015](Cm) coordinate[pos=1.015](Dm);
	\draw [thin] (Cm) -- ++(-0.8cm,0.8cm);
}
\path (A) -- (B) -- (D) -- (C);
\draw [thin] (D) -- ++(1.5cm,0) coordinate[pos=0.5](Dm); 
\draw [thin,<->] (Dm) -- ++(0,-10cm) node[pos=0.5,right] {$h_2$};
\draw [thin] (C) -- ++(-1.5cm,0) coordinate[pos=0.5](Cm); 
\draw [thin,<->] (Cm) -- ++(0,-6cm) node[pos=0.5,left] {$h_1$};
\draw [thin] (A) -- ++(0,-2cm) coordinate[pos=0.7](Am); 
\draw [thin] (B) -- ++(0,-2cm) coordinate[pos=0.7](Bm);
\draw [thin,<->] (Am) -- (Bm) node[pos=0.5,below] {$B$};
\draw [thick,<-] (B) ++(-2.8cm,0.8cm) -- ++(2.5cm,0) node[pos=.5,above]{$U$};
%
	\pgfresetboundingbox
	\path[opacity=0] (A) -- (D);
	\draw[opacity=0] (A) -- ++(0,-2cm) node[pos=0.7,below] {$B$};
	\path[coordinate] (A) -- (B) coordinate[pos=-0.3](Am) coordinate[pos=1.3](Bp);
	\draw[opacity=0] (Am) -- (Bp);
\end{tikzpicture}
}%
~\hspace{8ex}
\subfloat[Dimensionless]{%
\label{fig::dimensionless}%
        \centering
	\begin{tikzpicture}[scale=0.4]
	\footnotesize
\coordinate (A) at (0,0) {};
\coordinate (B) at (10cm,0) {};
\coordinate (C) at (0,7.5cm) {};
\coordinate (D) at (10cm,12.5cm);
\node at (5cm,5cm){$\hat{\Omega}$};

\draw [thick] (A) -- (B);
\draw [thick] (C) -- (D);
\draw  (A) -- (B) node[pos=0.5,above]{$\hat\Gamma_{\mathrm{plane}}$} %
		-- (D) node[pos=0.5,left]{$\hat\Gamma_{\mathrm{in}}$} %
		-- (C) node[pos=0.5,below=0.2cm]{$\hat\Gamma_{\mathrm{slider}}$} %
		-- (A) node[pos=0.5,right]{$\hat\Gamma_{\mathrm{out}}$};
\draw [->] (A) -- ++(12cm,0) node[right]{$\hat{x}$};
\draw [->] (A) -- ++(0,14cm) node[above]{$\hat{y}$};

\coordinate (Am) at (A);
\coordinate (Bm) at (B);
\foreach \x in {1,...,50}{
	\path[coordinate] (Am) -- (Bm) coordinate[pos=0.02](Am) coordinate[pos=1.02](Bm);
	\draw [ultra thin] (Am) -- ++(-0.8cm,-0.8cm);
}
\coordinate (Cm) at (C);
\coordinate (Dm) at (D);
\foreach \x in {1,...,66}{
	\path[coordinate] (Cm) -- (Dm) coordinate[pos=0.015](Cm) coordinate[pos=1.015](Dm);
	\draw [ultra thin] (Cm) -- ++(-0.8cm,0.8cm);
}
\pgfresetboundingbox
\path (A) -- (B) -- (D) -- (C);
\draw [ultra thin] (A) -- ++(-.8cm,0) node[left]{$0$}; 
\draw [ultra thin] (C) -- ++(-.8cm,0) node[left]{$\hat{h}_1$}; 
\draw [ultra thin, dashdotted] (C) ++(0,2.5cm) ++(5cm,0) -- ++(-5cm,0) -- ++(-.5cm,0) node[left]{$1$}; 
\draw [ultra thin] (C) ++(0,5cm) ++(10cm,0) -- ++(-10cm,0) -- ++(-.5cm,0) node[left]{$\hat{h}_2$}; 
\draw [ultra thin] (A) -- ++(0,-.5cm) node[below]{$0$}; 
\draw [ultra thin] (B) -- ++(0,-.5cm) node[below]{$1$};
\draw [thick,<-] (B) ++(-2.8cm,0.8cm) -- ++(2.5cm,0) node[pos=.5,above]{$\hat{U}=1$};
	\pgfresetboundingbox
	\path[opacity=0] (A) -- (D);
	\draw[opacity=0] (A) -- ++(0,-2cm) node[pos=0.7,below] {$B$};
	\draw[opacity=0] (A) -- ++(0,14cm) node[above]{$\hat{y}$};
\end{tikzpicture}
}%
\caption{Plane slider geometry}\label{fig:geometry}
\end{figure*}
%

\subsection{Plane slider geometry}
The geometry of the plane slider is illustrated in \Figure{:geometry}.
The rigid slider is fixed in the space 
above the horizontal plane which is moving in the horizontal direction
steadily with the speed~$U$. 
The lubricating fluid is dragged by the moving plane and forced through the converging gap.
The two solid surfaces define the natural boundaries 
$\Gamma_{\mathrm{plane}}$, $\Gamma_{\mathrm{slider}}$ 
of the domain $\Omega$, 
while the two artificial boundaries 
$\Gamma_{\mathrm{in}}$,  
$\Gamma_{\mathrm{out}}$
are defined at the inlet and outlet. 
The length of the domain is usually denoted by~$B$, 
and $h_1$, $h_2$ denote the height of the fluid film at 
the outlet and inlet, respectively.

A~crucial feature of the lubrication problem is that  $h=\tfrac12(h_1+h_2)$
is much smaller than $B$, that is 
$$	h = \eps B	,\qquad	\textrm{where } \eps \ll 1
.$$
We exploit this feature in the dimensionless formulation of the governing equations
in \Section{dimensionless}.
Note that apart from $B$, the geometry of the plane slider is characterized by 
two dimensionless parameters, $\eps$ and the ratio $h_2/h_1$.

\subsection{Boundary conditions}
\label{sec:bc}

We assume no-slip conditions at the solid walls,
i.e., that the velocity of the fluid vanishes on the upper surface,
while on the lower plate it equals the given tangential velocity
\begin{equation}\label{eq:Dirichlet}
	\vect{v}=\vect{0} \textrm{ on } {\Gamma}_{\mathrm{slider}}
\quad\textrm{and}\quad
	\vect{v}           =(-U,0) \textrm{ on } {\Gamma}_{\mathrm{plane}}
,\quad	U>0
.\end{equation}

The inflow and outflow boundaries are artificial,
subject to a~mass flux that is not known a~priori.
Therefore, there is no obvious proper choice for boundary conditions on $\Gamma_{\mathrm{in}}$ and $\Gamma_{\mathrm{out}}$.
A~detailed discussion of different possibilities is out of the scope of the present study.
For the moment, let us merely refer the reader to~\cite{HeywoodRannacherTurek1996},
appending the following remarks related to the particular situation in the plane slider.

When using the 
Reynolds approximation, one arrives at a~single equation for the pressure, the velocity being dealt with implicitly 
within the context of lubrication assumptions.
It is then straightforward to prescribe
\begin{equation}\label{eq:p0}
	p = P_0 \in \mathbb{R}
\quad	\textrm{at both the inlet and outlet}
,\end{equation}
where, most often, $P_0=0$ is chosen to represent the ambient pressure
(since that is presumably negligible compared to 
the pressure generated within the flow). 
 It is worth mentioning that for higher values of $\Rey_\eps$
(the modified Reynolds number as defined in \Section{dimensionlessform})
the inlet and outlet conditions for the Reynolds approximation should 
include the influence of the fluid inertia as well, see e.g.~\cite{Buckholz_1987}. 

In contrast, when the~weak solution to \eqref{eq:steady-incompressible} is considered,
the quantities naturally defined on the boundary are the vectors of velocity $\vect{v}$ and traction $-\tens{T}\vect{n}$
($\vect{n}$ denotes the outer normal vector).
Surprisingly, the boundary conditions on artificial boundaries seem to be an~issue that has not yet been unequivocally resolved 
in the literature.
Moreover, we should bear in mind two particular aspects of this study, namely: (a) that it should be possible to 
relate the problem setting to the Reynolds approximation approach, and (b) that we are keen to 
relate the variations of the pressure across the film to the piezoviscous response of the fluid.
To this end, we take advantage of the boundary condition which (a) results in the pressure values
being equal or approximately equal to the given constant $P_0$ and (b) does not induce cross flow and 
pressure variations in the vicinity of the artificial boundary.

Therefore, we prescribe 
\begin{equation}\label{eq:bc-artificial}
\begin{array}{l}
	- \tens{T}\vect{n} = 
	  \vect{b}_{\vect{n}} + \vect{b}_{\vect{\tau}}
\quad	\textrm{on $\Gamma_{\mathrm{in}}\cup\Gamma_{\mathrm{out}},$}
\\	\textrm{where}\quad
	  	\vect{b}_{\vect{n}} = P_0\,\vect{n}
\quad	\textrm{and}
\quad		\vect{b}_{\vect{\tau}}
		= \eta \left( \nabla\vect{v} - (\nabla\vect{v})^T \right) \vect{n}
.\end{array}
\end{equation}
Note that $\vect{b}_{\vect{\tau}}\cdot\vect{n}=0$.
Denoting $[\vect{w}]_{\vect{\tau}} \defeq \vect{w} - (\vect{w}\cdot\vect{n})\vect{n}$, 
one notices that $[\vect{b}_{\vect{n}}]_{\vect{\tau}}=0$.
The notation $\vect{b}_{\vect{n}}$ and $\vect{b}_{\vect{\tau}}$ thus corresponds to the 
decomposition of the prescribed traction into its normal and tangential parts.
We make the following observations concerning the above two terms.

\paragraph{First,}
the available theoretical results that guarantee the existence 
(and for small data, in certain sense, the uniqueness) of
the~weak solution to the system \eqref{eq:steady-incompressible} 
require, as one of the assumptions, that
\begin{equation}\label{eq:Tnbv}
\begin{array}{l}
	-\tens{T}\vect{n} = \vect{b}(\vect{v})
\quad	\textrm{on $\Gamma_{\mathrm{in}}\cup\Gamma_{\mathrm{out}},$}
\\	\textrm{where}
\quad	\vect{b}(\vect{v})\cdot\vect{v} \geq -\frac{\rho}{2} |\vect{v}|^2 (\vect{v}\cdot\vect{n})
	+ C
\end{array}
\end{equation}
is prescribed, where $C$ represents terms supposed to be of lower order in $\vect{v}$.
If~\eqref{eq:Tnbv} is not ensured then one cannot derive the standard energy estimates,
and a~weak solution with bounded kinetic energy is not necessarily found.
In the case of constant viscosity (i.e., for steady Navier--Stokes equations)
this is well known, see e.g.~\cite{BruneauFabrie1996,KracmarNeustupa01,Neustupa_2016};
the case of viscosity depending on pressure and shear rate is not different in this particular regard, see~\cite{LanzStebel2011}.
Note in particular, that \eqref{eq:Tnbv} does not allow one to prescribe the normal component of the traction 
independent of the velocity.
Neither \eqref{eq:bc-artificial}, nor the condition prescribing the constant traction, 
\begin{equation}\label{eq:Tnb0}
	-\tens{T}\vect{n} = P_0\vect{n}
,\end{equation}
nor, e.g., the boundary condition
\begin{equation}\label{eq:do-not}
	p\vect{n} - \eta \frac{\partial\vect{v}}{\partial\vect{n}} = P_0\vect{n}
,\end{equation}
are covered by~\eqref{eq:Tnbv}.
The latter is well known as the \emph{do-nothing} condition 
in the case that the viscosity is constant and 
that $P_0=0$.
In fact, in the case of a~radial flow, one can observe 
both the trivial and a~non-trivial solution for trivial boundary data, 
using any of the boundary conditions%
\footnote{
	To present such examples in detail would be out of the scope of this study 
	and is a~subject of a~work in preparation by J.~Hron and M.~Lanzend{\"o}rfer.
} %
\eqref{eq:bc-artificial}, \eqref{eq:Tnb0} or \eqref{eq:do-not}.
Even in the case of flow in straight channels or, importantly, the plane slider flow, 
one may indeed encounter complications in finding the numerical solution
(observing a~numerical blow-up of the kinetic energy in the approximate solution iterates).
However, such difficulties were not observed within the range of parameters presented in this paper.

Note that in case of pure outflow, $\vect{v}\cdot\vect{n}\geq0$,
the term that is cubic in the velocity in \eqref{eq:Tnbv} is negative 
and does not represent a~restriction on $\vect{b}$.
This is why \eqref{eq:Tnbv} does not restrain one to use \eqref{eq:Tnb0} or \eqref{eq:do-not}
in practical computations for outflow boundary conditions, as long as no backward flow is expected.
At the inflow, the velocity profile is then usually given explicitly as 
$$	\vect{v} = \vect{v}_{\mathrm{in}}
.$$
We remark that the above specification is not suitable for the plane slider problem and for most lubrication problems,
since~$\vect{v}_{\mathrm{in}}$ would not be known a~priori 
(not even the flux $\int_{\Gamma_{\mathrm{in}}}\vect{v}\cdot\vect{n}\,\ud\vect{s}$).

Note also that based on \eqref{eq:Tnbv} one could arrive at the idea of prescribing, e.g.,
$$
	\vect{b}_{\vect{n}}(\vect{v}) = 
		\left( 
		P_0  -  \frac{\rho}{2} |\vect{v}|^2 
		\right) \vect{n}
.$$
Such a~choice would lead to significant variations of the normal traction $-\tens{T}\vect{n}\cdot\vect{n}$
(and, consequently, of the resulting pressure) across the film 
and to a~concomitant cross-flow in the vicinity of both artificial boundaries,
even in the case of the flow between parallel plates.
In contrast, 
\eqref{eq:bc-artificial} gives the normal traction $-\tens{T}\vect{n}\cdot\vect{n} = \vect{b}_{\vect{n}}\cdot\vect{n} = P_0$ 
which is constant across the film and is satisfied by simple unidirectional flows.

\paragraph{Second,}
the particular relation for $\vect{b}_{\vect{\tau}}$ in~\eqref{eq:bc-artificial} 
was also chosen for the purpose of avoiding the pressure variations along the artificial boundary.
Indeed, the condition $\vect{b}_{\vect{\tau}}=\vect{0}$~\eqref{eq:Tnb0}
is not satisfied by simple unidirectional flows and it 
would result in the flow with the streamlines distorted 
and with the sharp pressure artifacts near the corners adjacent to the artificial boundary,
see the discussion and numerical examples in~\cite{HeywoodRannacherTurek1996}.
In contrast, 
with $\vect{b}_{\vect{\tau}}$ from \eqref{eq:bc-artificial} one can infer 
(formally, i.e.\ assuming that all the quantities are well defined on the boundary)
that
\begin{multline*}
	\vect{0}
=	[ \tens{T}\vect{n} + \vect{b}_{\vect{\tau}} ]_{\vect{\tau}} 
=	[ -p\,\vect{n} + \eta\,(\nabla\vect{v}+(\nabla\vect{v})^T)\vect{n}
\\
		       + \eta\,(\nabla\vect{v}-(\nabla\vect{v})^T)\vect{n} 
	 ]_{\vect{\tau}} 
=	2\eta [(\nabla\vect{v})\vect{n} ]_{\vect{\tau}}
\end{multline*}
implying, due to the viscosity being positive (while it need not be a~constant), that
$$	\left[ \frac{ \partial\vect{v} }{ \partial\vect{n} } \right]_{\vect{\tau}}  =
	\frac{ \partial [\vect{v}]_{\vect{\tau}} }{ \partial\vect{n} } = \vect{0}
.$$
This relation seems to have no physical interpretation 
except that, notably, it is satisfied by unidirectional flows perpendicular to the artificial boundary
(i.e., when $[\vect{v}]_{\vect{\tau}}\equiv \vect{0}$).
In other words, \eqref{eq:bc-artificial} does not induce cross-flow at the vicinity of inflow and outflow boundaries,
allowing thus for straight streamlines and the pressure field with no local artifacts in the corners.

Note that \eqref{eq:bc-artificial} can be formally rewritten as
$$	p\vect{n} - 2\eta \frac{\partial\vect{v}}{\partial\vect{n}} = P_0\vect{n}
,$$
a form similar to~\eqref{eq:do-not}.
For Navier--Stokes equations, due to the constraint of incompressibility $\divv\vect{v}=0$ 
and due to the viscosity being constant, the following
$\divv\tens{T} = \divv(-p\tens{I}+\eta(\nabla\vect{v})\,)$ holds.
If the weak formulation is based on this form involving the full velocity gradient,
then the \emph{do-nothing} boundary condition \eqref{eq:do-not} with $P_0=0$ 
corresponds to the trivial (zero) boundary term in the weak formulation,
see~\cite{HeywoodRannacherTurek1996} for details.
For fluids with variable viscosity, however, 
to define the weak solution based on the Cauchy stress tensor $\tens{T}$ 
and to give the boundary data in terms of the traction $-\tens{T}\vect{n}$
is more appropriate.
In this sense and in view of the previous paragraph, 
one can look on~\eqref{eq:bc-artificial} as a~generalization of the \mbox{\emph{do-nothing}} boundary condition
in the case of variable viscosity.

To our knowledge, there is no result concerning the existence of weak solutions to~\eqref{eq:steady-incompressible}
that would cover the presence of $\vect{b}_{\vect{\tau}}$ defined in~\eqref{eq:bc-artificial} in the boundary data.
The available theory is built upon uniform estimates for $\vect{v}$ in the 
Sobolev space $\WOd{1}{r}{}$, $1<r\leq2$,
and does not allow one to treat the gradients of velocity on the boundary.
Nevertheless, 
we did not encounter any complications related to $\vect{b}_{\vect{\tau}}$
in our numerical computations.

\subsection{The dimensionless formulation of \eqref{eq:steady-incompressible}}
\label{sec:dimensionlessform}
Let us rewrite the governing equations using the dimensionless variables
(indicated by hat).
Denote $\vect{x} = (x,y)$, 
$\vect{v} = (u,v)$ 
and $\hat{\vect{x}}$, $\hat{\vect{v}}$ analogously.
For simplicity, we neglect the body forces by assuming that $\vect{f}=\vect{0}$
and make use of that the fluid is homogeneous and incompressible by taking
$\rho = \rho^* \equiv \mathrm{const}$.
We define 
\begin{equation*}
\setlength{\arraycolsep}{2pt}
	\begin{array}{rcl}
		x &=& X^* \hat{x},
	\\	y &=& \eps X^* \hat{y},
	\end{array}
\quad
	\begin{array}{rcl}
		u &=& U^* \hat{u},
	\\	v &=& \eps U^* \hat{v},
	\end{array}
\qquad		
	p = P^*\hat{p},
	\quad
	\eta=\eta^*\hat{\eta}
.\end{equation*}
For the plane slider problem, 
we take $X^*=B$ and $U^*=U$ for the characteristic length and velocity.
As illustrated in \Figure{:dimensionless}, 
the plane slider geometry $\Omega$ transforms into the dimensionless 
$\hat{\Omega} = \{ (\hat{x},\hat{y}) \,:\; \hat{x}\in(0,1), \hat{y}\in(0,\hat{h}(\hat{x}))$,
where 
%
$\hat{h}(\hat{x})=\hat{h}_1+\hat{x}(\hat{h}_2-\hat{h}_1))$ and 
$\hat{h}_1 = \frac{2}{1+(h_2/h_1)}$, $\hat{h}_2 = \frac{2(h_2/h_1)}{1+(h_2/h_1)}$.
For more details see, e.g.,~\cite{Szeri_Lubrication_book1998}.

We set $\eta^*$ to be the viscosity at negligible shear rates and pressure
and define the characteristic pressure $P^*$
and the modified Reynolds number $\Rey_\eps$ 
(leaving $\Rey$ for the standard Reynolds number)
as customary by
\begin{equation*}
	P^* = \frac{ \eta^* U^* }{\eps^2 X^*}
\quad	\textrm{and}
\quad	\Rey_\eps = \eps \Rey = \frac{ \eps^2 \rho^* X^* U^*}{\eta^*}
	= \frac{ \rho^* {U^*}^2 }{P^*}
.\end{equation*}
Setting $\Rey_\eps=0$ 
represents Stokes-type flow, where the inertia of the fluid is neglected.
One easily rewrites \eqref{eq:steady-incompressible} as  
\begin{equation}\label{nondim:steady-incompressible}
\left.  \begin{array}{rcl}
	\divv_{\hat{\vect{x}}} \hat{\vect{v}} &=& 0
\\	\Rey_\eps
	\left(\begin{array}{c}
		\;\;\;   \hat{\vect{v}} \cdot \nabla_{\hat{\vect{x}}}{u}
	\\	\eps^2\, \hat{\vect{v}} \cdot \nabla_{\hat{\vect{x}}}{v}
	\end{array}\right)
	- \divv_{\hat{\vect{x}}} \widetilde{\tens{T}}
	& = & 
	\vect{0}
 \end{array}
\right\}
\quad	\textrm{in }\hat{\Omega},
\end{equation}
where
$	\widetilde{\tens{T}} = - \hat{p} \tens{I} + 2\hat{\eta}\widetilde{\tens{D}}_\eps
$ and
\begin{equation}\label{eq:Ttilde}
	\widetilde{\tens{D}}_\eps = \frac12\left(
		\begin{array}{cc}	
			2\eps^2 \partial_{\hat{x}}\hat{u}
		&	\eps^2 \partial_{\hat{x}}\hat{v} + \partial_{\hat{y}}\hat{u} 
		\\	\eps^4 \partial_{\hat{x}}\hat{v} + \eps^2 \partial_{\hat{y}}\hat{u} 
		&	2 \eps^2 \partial_{\hat{y}}\hat{v}
		\end{array}\right)
.\end{equation}
Note that $\widetilde{\tens{T}}$ differs from $\hat{\tens{T}}$ 
defined by
$	\tens{T} = 
	{P^*} \hat{\tens{T}}
$, wherein
$	\hat{\tens{T}} = - \hat{p} \tens{I} + 2\hat{\eta} \eps \hat{\tens{D}}_\eps
$ and
\begin{equation}\label{eq:Dhat}
	\hat{\tens{D}}_\eps = \frac{1}{2}\left(
		\begin{array}{cc}	
			2\eps \partial_{\hat{x}}\hat{u}
		&	\eps^2 \partial_{\hat{x}}\hat{v} +  \partial_{\hat{y}}\hat{u} 
		\\	\eps^2 \partial_{\hat{x}}\hat{v} +  \partial_{\hat{y}}\hat{u} 
		&	2 \eps \partial_{\hat{y}}\hat{v}
		\end{array}\right)
.\end{equation}

The no-slip boundary condition \eqref{eq:Dirichlet}
takes the simple form
\label{sec:dimensionless}
\begin{equation}\label{nondim:Dirichlet}
	\hat{\vect{v}}=\vect{0} \textrm{ on } \hat{\Gamma}_{\mathrm{slider}}
\quad\textrm{and}\quad
	\hat{\vect{v}}                       =(-1,0) \textrm{ on } \hat{\Gamma}_{\mathrm{plane}}
.\end{equation}
Following~\eqref{eq:Ttilde}, one can easily derive %
(here we take the advantage of that the artificial boundary is 
perpendicular to the $x$-axis, so that $\vect{n}=\pm(1,0)=\hat{\vect{n}}$ holds)
that \eqref{eq:bc-artificial} takes the dimensionless form 
\begin{equation}\label{nondim:bc-artificial}
	- \widetilde{\tens{T}} \hat{\vect{n}} 
=	\hat{\vect{b}}
\defeq  
	\left\{	\begin{array}{rl}
		(\, 	\hat{P}_0 ,\, \hat{b}_\tau 
		\,)
	&	\qquad	\textrm{on $\hat{\Gamma}_{\mathrm{in}}$}
	,\\ -	(\, 	\hat{P}_0 ,\, \hat{b}_\tau 
		\,)
	&	\qquad	\textrm{on $\hat{\Gamma}_{\mathrm{out}}$}
,	\end{array}\right.
\end{equation}
where $\hat{P}_0 = P_0 / P^*$
and $\hat{b}_\tau = \hat{\eta}\, ( \eps^4 \partial_{\hat{x}}\hat{v} - \eps^2 \partial_{\hat{y}}\hat{u} )$.
Note in particular, that \eqref{eq:bc-artificial} reduces formally to \eqref{eq:p0}
when the lubrication assumptions are taken,
namely when $\partial[\vect{v}]_{\vect{\tau}}/\partial\vect{n}$ 
and $\partial(\vect{v}\cdot\vect{n})/\partial\vect{n}$ can be neglected.

\subsection{Viscosity}
We are interested in lubrication problems wherein the range of pressures involved
is very large and in virtue of which the viscosity of the fluid changes by several 
orders of magnitude, in fact by as much as $10^6$ or $10^8$.
That this is indeed the case is borne out by experiments. 
It is also well known that many lubricants shear-thin and thus we employ the model
wherein the viscosity depends on both the pressure and the shear rate 
(in the general three dimensional or planar flow on the norm of the symmetric part of the velocity gradient).
Several correlations have been used to describe the variation of the viscosity with pressure.
In this study we will follow the model suggested 
by Bair \cite{Bair2006}
where the viscosity is related to the pressure and to the Frobenius norm of the velocity 
gradient through the Carreau--Yasuda relationship.
We shall specifically assume that the viscosity is given by the following relation,
with $\eta_0>0$, $1<r<2$,
\begin{equation}\label{eq:CY}
	\eta = \eta_0 \, a(p) \left( 1 + b(p) \tr\tens{D}^2 \right)^{\frac{r-2}{2}}
,\end{equation}
where 
$a(\cdot)$, $b(\cdot)$ are given functions%
\footnote{ 
The three reference lubricants presented 
by Bair \cite{Bair2006}
are characterized as compressible, their viscous response depending on the density and temperature,
$$	\eta \equiv \eta_{\mathrm{com}}(\rho, \tr\tens{D}^2, \vartheta)
.$$
In view of incompressibility and the assumption of isothermal conditions,
we consider the pressure and shear-rate dependent viscosity only, i.e.
$$	\eta \defeq \eta(p, \tr\tens{D}^2) 
=	\eta_{\mathrm{com}}( \rho_{\mathrm{com}}(p,\underline{\theta}), \tr\tens{D}^2, \underline{\theta} )
,$$
    where the material properties are considered at constant temperature 
and where the density $\rho_{\mathrm{com}}(p,\underline{\theta})$ 
merely provides the correct dependence of the viscosity on the pressure, 
the actual density considered in the momentum conservation being 
constant.
} 
of the pressure~$p$.
In~order to simplify the discussion of the numerical results in the dimensionless formulation of the problem,
we take the idealized exponential model for the pressure--viscosity dependence $a(p)$
and an~analogous simple relation for the shifting rule $b(p)$, i.e., 
with $\eta_0, G, \alpha, \beta > 0$, $1<r<2$,
\begin{equation}\label{eq:expCY}
	\eta = \eta_0 e^{\alpha p} \left( 1 + G e^{\beta p} \tr\tens{D}^2 \right)^{\frac{r-2}{2}}
.\end{equation}
Note that for small shear rates, \eqref{eq:expCY} reduces to 
$$	\eta \sim \eta_0 e^{\alpha p}
	\qquad\textrm{(for $\tr\tens{D}^2 \ll e^{-\beta p}/G$)}
,$$ 
while for large shear rates there is, with $\tilde{\beta} = \alpha-\tfrac{2-r}{2}\beta$,
$$	\eta \sim \eta_0 e^{\tilde{\beta}p} \sqrt{G\tr\tens{D}^2}^{\,r-2}
	\qquad \textrm{(for $\tr\tens{D}^2 \gg e^{-\beta p}/G$)}
.$$
Finally, it follows from the definition of $\hat{\tens{D}}_\eps$ from \eqref{eq:Dhat} that%
\footnote{
Note that $D^*$ represents the characteristic shear rate.
Note also that $\tr\hat{\tens{D}}_\eps^2 \sim \tfrac12(\partial_{\hat{y}}\hat{u})^2$, as $\eps\searrow0$.
} 
$$	\tr \tens{D}^2 = {D^*}^2 \tr \hat{\tens{D}}_\eps^2
,	\qquad\textrm{where }
	D^* = \frac{U^*}{\eps X^*}
,$$
whereby we obtain the following dimensionless form of \eqref{eq:expCY}, %
\begin{equation}\label{nondim:expCY}
	\hat{\eta} = e^{\alpha^* \hat{p}} 
	\left(  
		1 + G^* e^{\beta^* \hat{p}} \tr \hat{\tens{D}}_\eps^2 
	\right)^{\frac{r-2}{2}}
,\end{equation}
provided that
\begin{equation*}
	\alpha^* = \alpha P^*
,\quad	\beta^* = \beta P^* 
,\quad	G^* = G {D^*}^2
\quad	\textrm{and}\quad
	\eta^* = \eta_0
.\end{equation*}
The numerical simulations presented in what follows will be restricted,
for the sake of simplicity,
to $r=3/2$ and $\beta/\alpha=\beta^*/\alpha^*=2$,
leaving two remaining parameters: $\alpha^*$ and $G^*$.

\section{Numerical solution}
\label{sec:numerics}

We approximate the problem described by \eqref{nondim:steady-incompressible},
\eqref{eq:Ttilde}, \eqref{nondim:Dirichlet}, \eqref{nondim:bc-artificial} 
and~\eqref{nondim:expCY}
using the following Galerkin formulation:
Find $(\hat{\vect{v}}_l,\hat{p}_l) \in (\hat{\vect{v}}_0+\vect{X}_l)\times\mathcal{Q}_l$
(the discrete solution) such that
\begin{eqnarray}
	\int_{\hat\Omega} ( \divv_{\hat{\vect{x}}} \hat{\vect{v}}_l ) q \,\mathrm{d}\hat{x}
&=&	0 	
\qquad \forall q \in \mathcal{Q}_l
,\\	
	\Rey_\eps \, \int_{\hat\Omega}
	\left(\begin{array}{c}
		\;\;\;   \hat{\vect{v}}_l \cdot \nabla_{\hat{\vect{x}}}{u}_l
	\\	\eps^2\, \hat{\vect{v}}_l \cdot \nabla_{\hat{\vect{x}}}{v}_l
	\end{array}\right) \!\!
	\cdot \vect{w} \,\mathrm{d}\hat{x}
+	\int_{\hat\Omega} \widetilde{\tens{T}}_l \cdot \nabla_{\hat{\vect{x}}} \vect{w} \,\mathrm{d}\hat{x}
\hspace{-9ex}\nonumber&&\\
+	\int_{\hat\Gamma_{\mathrm{in}} \cup \hat\Gamma_{\mathrm{out}}}
	\hat{\vect{b}}_l \cdot \vect{w} \,\mathrm{d}\hat{s}
&=&	0
\qquad \forall \vect{w} \in \vect{X}_l
,\end{eqnarray}
with $\widetilde{\tens{T}}_l$, $\hat{\vect{b}}_l$ given by \eqref{eq:Ttilde}, \eqref{nondim:bc-artificial} 
and \eqref{nondim:expCY}.
The parameter $l>0$ is related to the 
finite-dimensional function spaces $\mathcal{Q}_l$, $\vect{X}_l$, 
\begin{align*}
&	\mathcal{Q}_l
\subset	L^{1}(\hat\Omega)
\qquad\textrm{and}\qquad
\\&	\vect{X}_l 
\subset	\left\{ \vect{w}\in W^{1,1}(\hat\Omega)^2 \,;\; \vect{w}=\vect{0}
	\textrm{ on } \hat\Gamma_{\mathrm{slider}}\cup\hat\Gamma_{\mathrm{plane}}
	\right\}
\end{align*}
and $\hat{\vect{v}}_0$ is a~suitable extension of the Dirichlet data \eqref{nondim:Dirichlet}.
Naturally, $\vect{X}_l$, $\mathcal{Q}_l$ 
are to be chosen such that all the integrals are well defined and finite.

The numerical simulations presented in this work are based on the following 
finite element approach.
The domain $\hat\Omega$ is discretized by means of quadrilaterals
(of diameter $l$ at most)
and $\vect{X}_l$, $\mathcal{Q}_l$ are generated by the second order 
$\mathbb{Q}_2/\mathbb{P}_{-1}$ finite element pair described in 
\cite{bookGreshoSani_2000ii, SaniEtal_1981} (the conforming biquadratic elements for the velocity
and the discontinuous piecewise linear space for the pressure).
The resulting system of nonlinear algebraic equations is solved using the damped
Newton method with line search, with the Jacobian matrix approximated by 
the central differences.
The linear subproblems, sparse and unsymmetric, are mostly solved 
by the direct sparse {LU} factorization 
implemented in the {UMFPACK} package, see~\cite{UMFPACK}.
The presented numerical simulations are performed on a~regular mesh 
of $3\cdot4^6$ finite elements, corresponding $136\,194$ degrees of freedom.

In an~ideal situation, letting the discretization parameter $l\searrow0$ 
and hence the dimension of the finite element function spaces $\vect{X}_l$, $\mathcal{Q}_l$ to infinity,
the error due to discretization would vanish
and the discrete solution $(\vect{v}_l,p_l)$ would eventually converge 
to a~(weak) solution $(\vect{v},p)$.
This desired behaviour has been guaranteed rigorously 
in~\cite{Hirn2012} after making additional requirements
which, however, do not cover realistic viscosity \eqref{eq:expCY} 
at large pressures.
The result in~\cite{Hirn2012} stems from intensive research devoted to 
the notion and existence of a~weak solution for 
incompressible fluids with pressure- and shear rate- dependent viscosity,
see~\cite{%
BulicekMalekRaj2009a,%
BulicekMalekRaj2009b,%
FrantaMalekRaj_05,%
Lanz_09%
} (see also \cite{%
JaneckaPrusa2014,%
RehorPrusa2016,%
HronMalekPrusaRaj_2009,%
BulicekMajdoubMalek_2010%
} and the references therein).
One of the assumptions embodied in the theoretical framework 
requires in particular that
\begin{equation}\label{eq:dSdp}
	\left|\left| \frac{\partial\tens{S}}{\partial p} \right|\right|
\leq	C
\leq	1,
\qquad	\textrm{where $\tens{S}=\tens{T}+p\tens{I}=2\eta\tens{D}$} 
,\end{equation}
for certain constant $C$,
see the concerned results for details
\footnote{
	One of the key steps when proving the existence of a~weak solution,
	to put it in a~simple way, 
	is to establish the uniqueness of the pressure field $p$ provided that the velocity field of the solution $\vect{u}$ is given.
	Depending on the setting of the problem
	(which includes a~number of assumptions concerning the domain geometry, the boundary conditions given, the parameters of the rheology, etc.)
	one should be able to obtain the inequality of the following type
	$$	0 < C < \inf_{q\in\mathcal{Q}} \sup_{\vect{\psi}\in\vect{X}}
		\frac{ \int_\Omega q \, \divv\vect{\psi} \,\mathrm{d}x }{ ||q||_2 \, ||\nabla\vect{\psi}||_2 }
	,$$
	where the functional spaces (and the corresponding norms in the above inequality) for the pressure and velocity, $\mathcal{Q}$ and $\vect{X}$,
	and the constant $0<C\leq1$ would depend on the particular setting.
	Here let us say $\mathcal{Q}\subset L^2(\Omega)$ and $\vect{X}\subset\{ \vect{\psi}\,;\; \nabla\vect{\psi}\in L^2(\Omega) \}$.

	With help of the above inequality and using the weak momentum equation, one can estimate for 
	two pressure fields $p_1$, $p_2$ and the given velocity field $\vect{u}$ 
	that the following 
	\begin{multline*}
		C || p_1 - p_2 ||_2 
	\leq	\sup_{ ||\nabla\vect{\psi}||_2 = 1 } \int_\Omega (p_1-p_2) \divv\vect{\psi} \,\mathrm{d}x
	\\=	\sup_{ ||\nabla\vect{\psi}||_2 = 1 } \int_\Omega \left( \tens{S}(p_1,\tens{D}) - \tens{S}(p_2,\tens{D}) \right)\cdot\nabla\vect{\psi} \,\mathrm{d}x
	\end{multline*}
	holds.
	One obtains the result by estimating the last term by
	\begin{multline*}	
		... 
		\leq	|| \tens{S}(p_1,\tens{D}) - \tens{S}(p_2,\tens{D}) ||_2
	\\	\leq	\left|\left| |p_1-p_2| \int_{p_1}^{p_2} 
				\frac{ \partial\tens{S}( p_1 + s(p_2-p_1) ,\tens{D}) }{ \partial p } \,\mathrm{d}s 
			\right|\right|_2
		<	C ||p_1-p_2||_2
	,\end{multline*}
	provided that $|\partial\tens{S}/\partial p|<C$.	
}%
.
Note that \eqref{eq:expCY} with $\alpha>0$ 
violates \eqref{eq:dSdp} {\em both} at elevated pressures
{\em or} high shear rates.
For \eqref{eq:expCY}, the~notion of a~solution 
such that the problem would be well posed
remain a~challenging open problem,
as far as no a~priori restrictions on the data size are imposed.

On the basis of our numerical computations, including those presented 
in the next section, 
\eqref{eq:dSdp} seems to be both sufficient and necessary
(with $C=1$, or nearly so) for the presented numerical approach to converge successfully.
Once \eqref{eq:dSdp} is violated by the approximate solution at hand,
we were unable to obtain any discrete solution.
An analogous restriction seems to apply for previously published results 
in a~more complex setting
as well, cf.~\cite{Knauf2013, Almqvist2002}. 

For the sake of completeness we recall that 
there are no theoretical well-posedness results allowing for the boundary condition \eqref{eq:bc-artificial},
as discussed already in \Section{bc}, cf.~\cite{LanzStebel2011}.
Note also that some lower values of the parameter $1<r\leq2$
are excluded in the well-posedness analysis, depending on the 
particular setting of the problem (see the above mentioned references).

\section{Numerical results}
\label{sec:results}
\subsection{Constant viscosity, $\Rey_\eps\geq0$}
With $\alpha^*=0$ and $G^*=0$ (or $r=2$) in \eqref{nondim:expCY}, 
the model reverts to that of an~incompressible Navier--Stokes fluid.
The non-dimensional plane slider flow problem is then described by the 
three parameters
$$	h_2/h_1, \qquad \eps \qquad\textrm{and}\qquad \Rey_\eps
$$
and by the pressure drop (the difference of the constants $\hat{P}_0$ in \eqref{nondim:bc-artificial} 
on $\Gamma_\mathrm{in}$ and $\Gamma_\mathrm{out}$).
We prescribe $\hat{P}_0=0$ on the both boundaries throughout the paper;
this represents the ambient pressure, 
supposedly negligible 
in comparison to the characteristic pressure $P^*$.
It is for the sake of simplicity that we keep $\hat{P}_0=0$ even for $\Rey_\eps>0$, 
cf.~\cite{Buckholz_1987}.

\begin{figure*}
\centering
\subfloat[$\hat{u}$]{
	\includegraphics[width=0.45\textwidth,trim=200 300 20 250,clip=true]{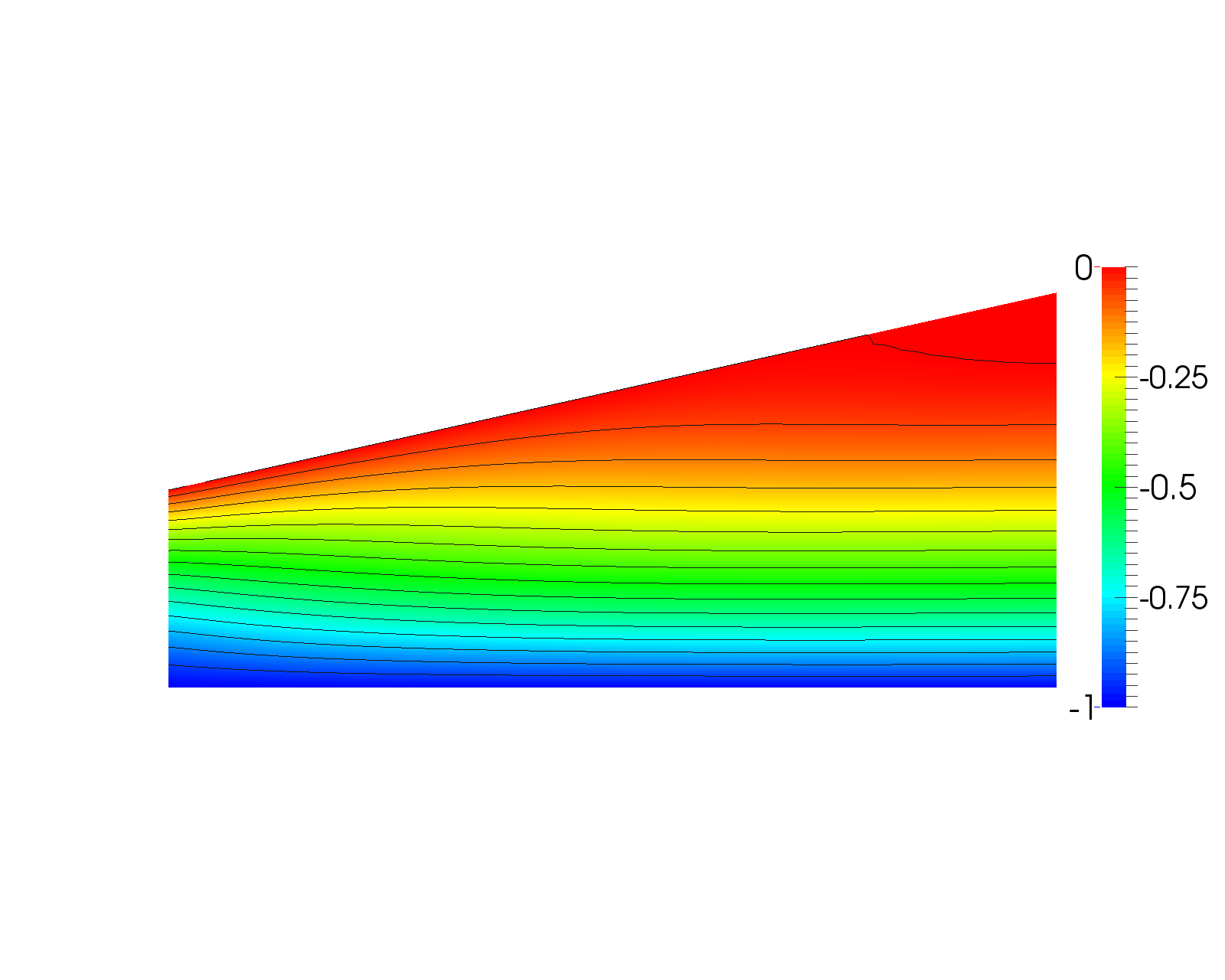} 
}
\hspace{2ex}
\subfloat[$\hat{v}$]{
	\includegraphics[width=0.45\textwidth,trim=200 300 20 250,clip=true]{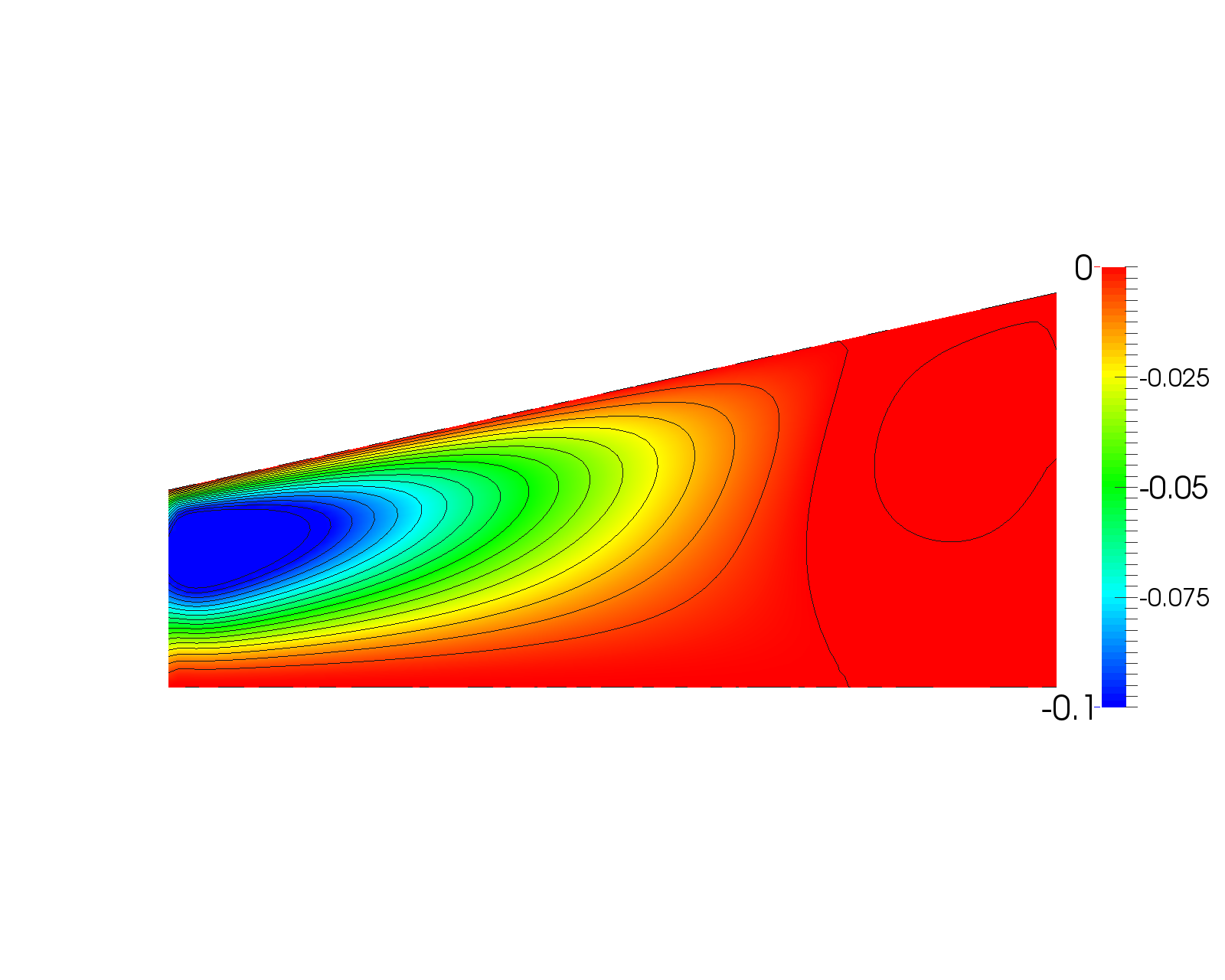} 
}
\\ 
\subfloat[$\hat{p}$]{
	\includegraphics[width=0.45\textwidth,trim=200 300 20 250,clip=true]{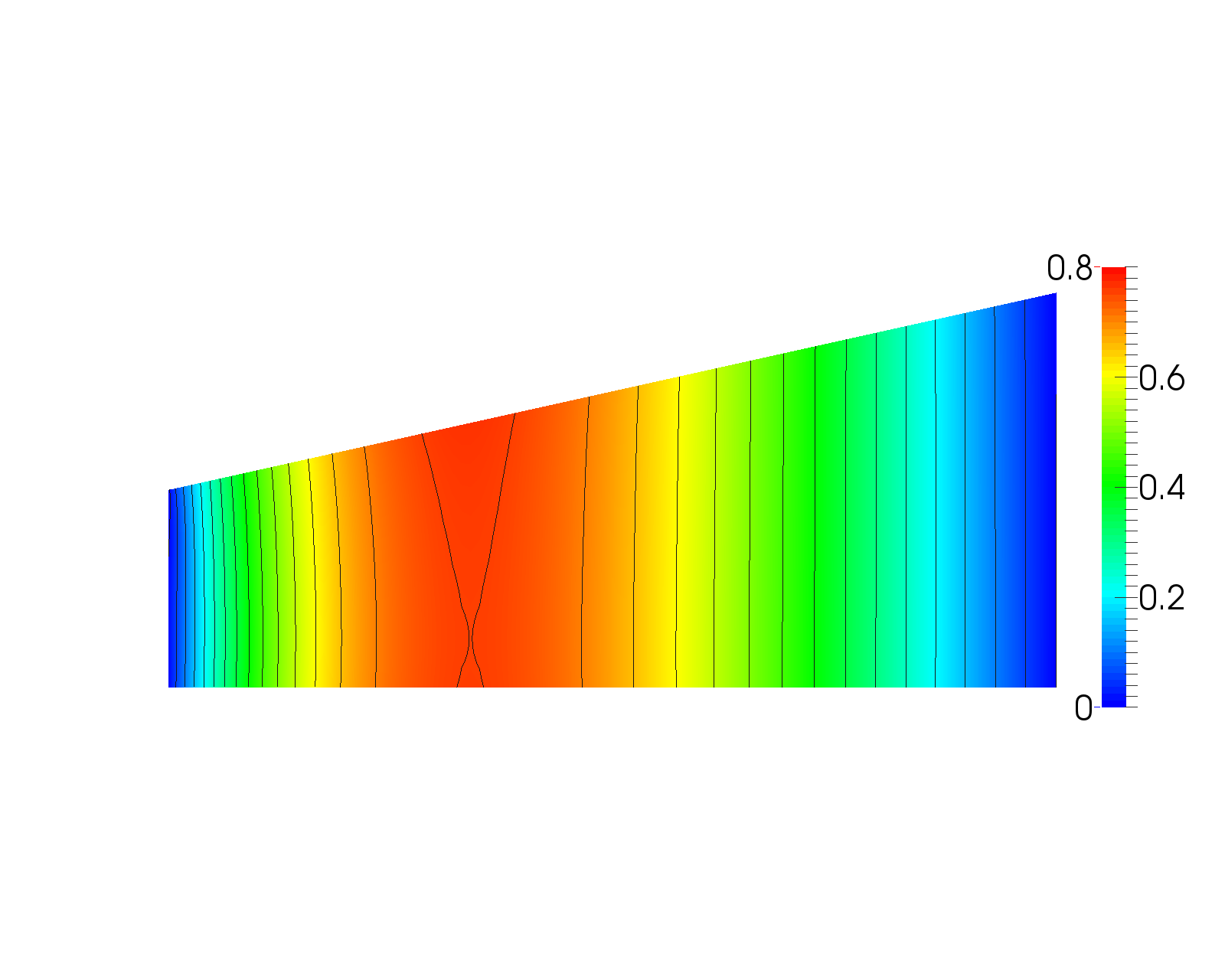} 
}
\hspace{2ex}
\subfloat[$\hat{p}$ for $\eps=0.005$]{
	\includegraphics[width=0.45\textwidth,trim=200 300 20 250,clip=true]{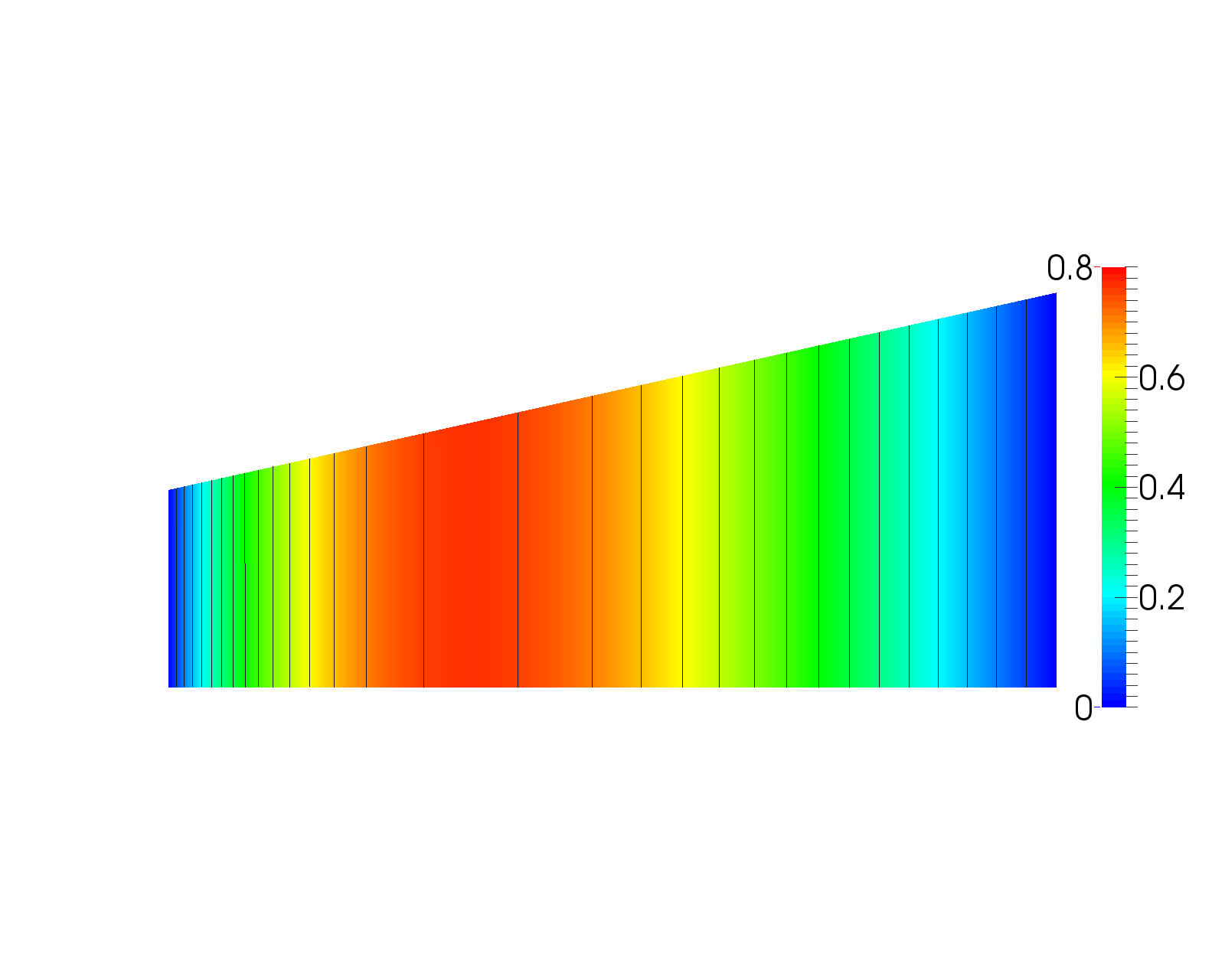} 
	\label{fig:Rey0-d}
}
\caption{ Flow in a~slider bearing
	($h_2/h_1=2$, $\Rey_\eps=10$, (a,b,c) $\eps=0.1$, (d) $\eps=0.005$)}
\label{fig:Rey0}
\end{figure*}

The resulting flow has a~rather simple structure, as illustrated in \Figure{Rey0}
for $h_2/h_1=2$, $\Rey_\eps=10$ and $\eps=0.1$.
The velocity field is not far from being unidirectional, its horizontal component $\hat{u}$ having a~parabolic profile
across the film.
A~pressure peak is generated in the center part of the domain.
The pressure differences across the film 
vanish for small values of $\eps$, as shown in \Figure{Rey0-d} for $\eps=0.005$.

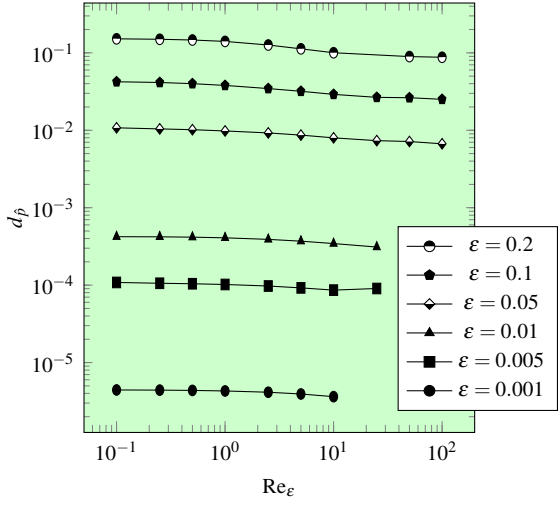
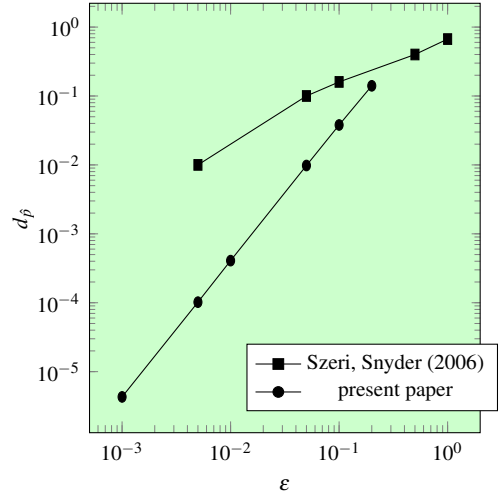
\begin{figure*}
\centering
\subfloat[$d_{\hat{p}}$ for various values of $\eps$ and $\Rey_\eps$]{%
\begin{minipage}{0.47\textwidth}
	\begin{tikzpicture}
	\vspace{1em}%
		\begin{loglogaxis}[
			xlabel=$\Rey_\eps$,  
			ylabel=$d_{\hat{p}}$,
			unbounded coords=discard,
			scale=1.0, xscale=0.75,
			font=\footnotesize,
			axis background/.style={fill=green!20},	
			legend style={anchor=south west,at={(0.95,0.05)},reverse legend,font=\footnotesize}, 
			cycle list name=mark list* 
		]
		\addplot+ [solid] table[x index=0,y index=2, smooth, solid, mark=.] {\dataDiffMine};
		\addplot+ [solid] table[x index=0,y index=3, smooth, solid, mark=.] {\dataDiffMine};
		\addplot+ [solid] table[x index=0,y index=4, smooth, solid, mark=.] {\dataDiffMine};
		\addplot+ [solid] table[x index=0,y index=5, smooth, solid, mark=.] {\dataDiffMine};
		\addplot+ [solid] table[x index=0,y index=6, smooth, solid, mark=.] {\dataDiffMine};
		\addplot+ [solid] table[x index=0,y index=7, smooth, solid, mark=.] {\dataDiffMine};
		\legend{ 
			{ $\eps=0.001$ },
			{ $\eps=0.005$ },
			{ $\eps=0.01$ },
			{ $\eps=0.05$ },
			{ $\eps=0.1$ },
			{ $\eps=0.2$ }
		}
		\end{loglogaxis}
	\end{tikzpicture}
	\label{plot:Dp--a}
\end{minipage}}
\hspace{2em}%
\subfloat[$d_{\hat{p}}$ variation with $\eps$, comparison to~\cite{SzeriSnyder_2006}]{%
\begin{minipage}{0.47\textwidth}
	\begin{tikzpicture}
		\begin{loglogaxis}[
			xlabel=$\eps$,  
			ylabel=$d_{\hat{p}}$,
			unbounded coords=discard,
			scale=1.0, xscale=0.75,
			font=\footnotesize,
			axis background/.style={fill=green!20},	
			legend style={anchor=south west,at={(0.23,0.05)},reverse legend}, 
			ytick={1e-5,1e-4,1e-3,1e-2,1e-1,1e0},
			cycle list name=mark list* 
		]
		\addplot+ [solid] table[x index=0,y index=1, smooth, solid, mark=.] {\dataDiffEpsCompare};
		\addplot+ [solid] table[x index=0,y index=2, smooth, solid, mark=.] {\dataDiffEpsCompare};
		\legend{ 
			{ present paper },
			{ Szeri, Snyder (2006) }
		}
		\end{loglogaxis}		
	\end{tikzpicture}
	\label{plot:DpComparison}
\end{minipage}}
\caption{Dimensionless pressure difference coeficient $d_{\hat{p}}$ 
	($h_2/h_1=2$ and 
	$\hat\eta\equiv1$)}
\label{plot:Dp}
\end{figure*}

\begin{figure*}
\centering
\subfloat[vertical component, $-\widetilde{\tens{T}}\hat{\vect{n}}\cdot\vect{e}_y$]{%
\begin{minipage}{0.47\textwidth}
	\begin{tikzpicture}
	\vspace{1em}%
		\begin{axis}[
			xlabel={ $\hat{x}$ },  
			ylabel={ $-\widetilde{\tens{T}}\hat{\vect{n}}\cdot\vect{e}_y$ },
			width={ 0.95\textwidth }, height={ 0.4\textheight },
			every axis y label/.style=
				{at={(ticklabel cs:0.7)},rotate=90,anchor=near ticklabel},
			unbounded coords=discard,
			font=\small,
			axis background/.style={fill=green!20},	
			legend pos=outer north east,
			cycle list name=linestyles*, 
			legend to name=null
		]
		\addplot+ [solid]  table[x index=0,y index=3, smooth, solid, mark=none,col sep=space,trim cells=true] {results/03002m6.Tnxy-slider.0001};
		  \addlegendentry{ $\Rey_\eps=0$, $\eps=0.1$ }
		\addplot+ [dotted] table[x index=0,y index=3, smooth, solid, mark=none,col sep=space,trim cells=true] {results/03005m6.Tnxy-slider.0001};
		  \addlegendentry{ $\Rey_\eps=0$, $\eps=0.005$ }
		\addplot+ [dashed]  table[x index=0,y index=3, smooth, solid, mark=none,col sep=space,trim cells=true]  {results/03052m6.Tnxy-slider.0001};
		  \addlegendentry{ $\Rey_\eps=2.5$, $\eps=0.1$ }
		\addplot+ [dotted] table[x index=0,y index=3, smooth, solid, mark=none,col sep=space,trim cells=true] {results/03055m6.Tnxy-slider.0001};
		  \addlegendentry{ $\Rey_\eps=2.5$, $\eps=0.005$ }
		\addplot+ [dashdotted]  table[x index=0,y index=3, smooth, solid, mark=none,col sep=space,trim cells=true]  {results/03072m6.Tnxy-slider.0001};
		  \addlegendentry{ $\Rey_\eps=10$, $\eps=0.1$ }
		\addplot+ [dotted] table[x index=0,y index=3, smooth, solid, mark=none,col sep=space,trim cells=true] {results/03075m6.Tnxy-slider.0001};
		  \addlegendentry{ $\Rey_\eps=10$, $\eps=0.005$ }	
		\end{axis}
	\end{tikzpicture}
\end{minipage}}
\hspace{0.03\textwidth}
\subfloat[horizontal component, $-\widetilde{\tens{T}}\hat{\vect{n}}\cdot\vect{e}_x$]{%
\begin{minipage}{0.47\textwidth}
	\centering
	\begin{tikzpicture}
	\vspace{1em}%
		\begin{axis}[
			xlabel={ $\hat{x}$ },  
			ylabel={ $-\widetilde{\tens{T}}\hat{\vect{n}}\cdot\vect{e}_x$ },
			width={ 0.95\textwidth }, height={ 0.4\textheight },
			every axis y label/.style=
				{at={(ticklabel cs:0.7)},rotate=90,anchor=near ticklabel},
			unbounded coords=discard,
			font=\small,
			axis background/.style={fill=green!20},	
			legend pos=south east,
			legend style={font=\footnotesize},
			cycle list name=linestyles*, 
		]
		\addplot+ [solid]              table[x index=0,y index=2, smooth, solid, mark=none,col sep=space,trim cells=true] {results/03002m6.Tnxy-slider.0001};
			\addlegendentry{ $\Rey_\eps=0$, $\eps=0.1$ }
		\addplot+ [dotted]             table[x index=0,y index=2, smooth, solid, mark=none,col sep=space,trim cells=true] {results/03005m6.Tnxy-slider.0001};
			\addlegendentry{ $\Rey_\eps=0$, $\eps=0.005$ }
		\addplot+ [dashed]              table[x index=0,y index=2, smooth, solid, mark=none,col sep=space,trim cells=true]  {results/03052m6.Tnxy-slider.0001};
			\addlegendentry{ $\Rey_\eps=2.5$, $\eps=0.1$ }
		\addplot+ [dotted]             table[x index=0,y index=2, smooth, solid, mark=none,col sep=space,trim cells=true] {results/03055m6.Tnxy-slider.0001};
			\addlegendentry{ $\Rey_\eps=2.5$, $\eps=0.005$ }
		\addplot+ [dashdotted]              table[x index=0,y index=2, smooth, solid, mark=none,col sep=space,trim cells=true]  {results/03072m6.Tnxy-slider.0001};
			\addlegendentry{ $\Rey_\eps=10$, $\eps=0.1$ }
		\addplot+ [dotted]             table[x index=0,y index=2, smooth, solid, mark=none,col sep=space,trim cells=true] {results/03075m6.Tnxy-slider.0001};
			\addlegendentry{ $\Rey_\eps=10$, $\eps=0.005$ }
		\end{axis}
	\end{tikzpicture}
\end{minipage}}
\caption{Dimensionless traction vector $-\widetilde{\tens{T}}\hat{\vect{n}}$ along the slider surface $\hat{\Gamma}_{\mathrm{slider}}$
	($h_2/h_1=2$ and 
			$\hat\eta\equiv1$)}
\label{plot:draglift}
\end{figure*}
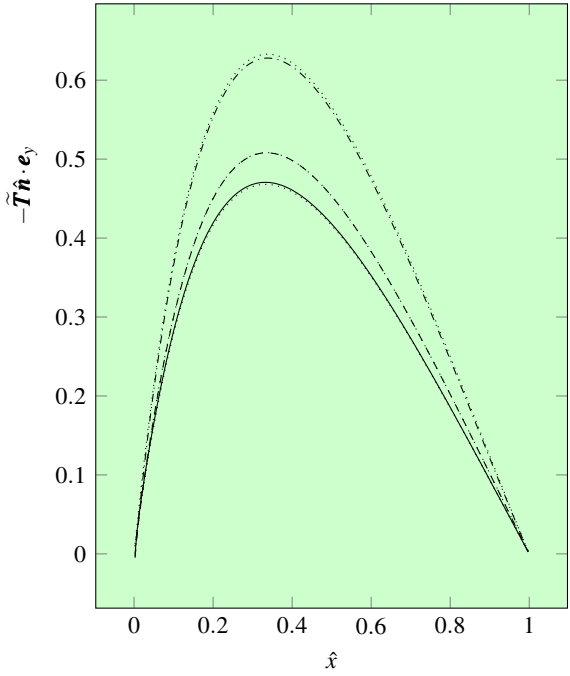
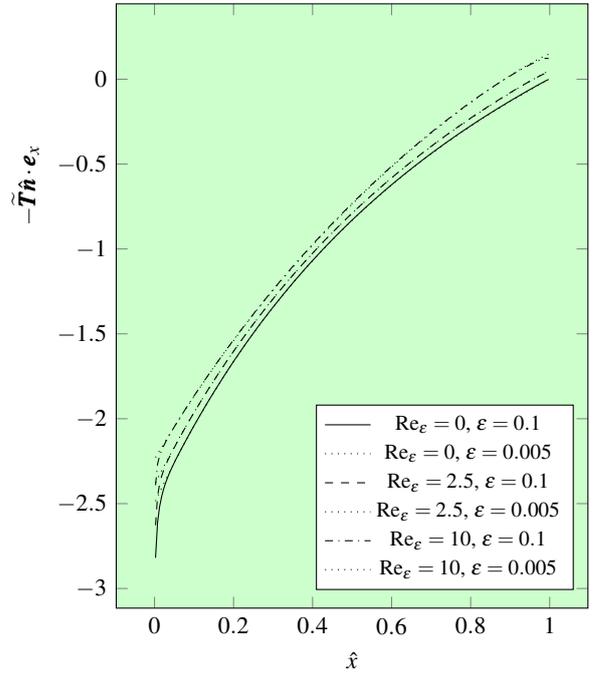

\begin{table}
\centering
\begin{tabular}{llll}
\midrule
	$h_2/h_1$	& $\eps=0.1$	& $\eps=0.01$	& $\eps=0.001$
\\
\midrule
       11.0     &  1.603          &  1.580          &   1.579     
\\     3.00     &  0.5965         &  0.5917         &   0.5917     
\\     2.25     &  0.4257         &  0.4229         &   0.4228     
\\     2.00     &  0.3597         &  0.3575         &   0.3575     
\\     1.50     &  0.2060         &  0.2050         &   0.2049     
\\     1.20     &  0.09181        &  0.09137        &   0.09136     
\\     1.10     &  0.04791        &  0.04769        &   0.04768     
\\     1.01     &  0.004999       &  0.004975       &   0.004975    
\\
\midrule
\end{tabular}
\caption{Dimensionless lift $\hat{F}_y$, for different $h_2/h_1$ and $\eps$ ($\hat{\eta}\equiv1$, $\Rey_\eps=0$)}
\label{table:lift}
\end{table}
The problem has been studied by Szeri and Snyder~\cite{SzeriSnyder_2006}, 
where the results obtained using the Reynolds lubrication approximation 
and the numerical results for a~quasi two-dimen\-sional thin-film flow model derived in the paper
were compared to the finite element solution to the full Navier--Stokes problem. 
The pressure differences across the film, quantified for convenience of the presentation by 
$$	d_{\hat{p}} = \frac{ \max_{\hat{x}\in(0,1)} | \hat{p}(\hat{x},\hat{h})-\hat{p}(\hat{x},0) | }{
		\max_{\hat{x}\in(0,1)} \hat{p}(\hat{x},\hat{h}) }
,$$
were computed for the Navier--Stokes solutions, for a~reasonable range of parameters,
$\Rey_\eps$ up to $100$ and $\eps$ from $0.005$ up to $1$.
It was observed that $d_{\hat{p}}$ does not increase with $\Rey_\eps$ and that it remains
small even for $\eps$ rather large.
Similarly, the dimensionless pressure peak, $\max_{\hat{\vect{x}}\in\hat{\Omega}} \hat{p}(\hat{\vect{x}})$,
same as the dimensionless force (lift) 
$\hat{F}_y$, where%
\footnote{
Note that $\ud s = X^* \sqrt{ \hat{n}_y^2 + \eps^2 \hat{n}_x^2 }\; \ud\hat{s}$.
}  
$$
	\hat{\vect{F}} 
=	\left(\begin{array}{c}  \hat{F}_x  \\  \hat{F}_y  \end{array}\right)
=	\int_{\hat{\Gamma}_{\mathrm{slider}}} 
	\hspace{-2ex}		- \widetilde{\tens{T}}\hat{\vect{n}} 	\,\ud \hat{s}
,\qquad	\textrm{where }
	\vect{F} 
=	P^* X^* 
	\left(\begin{array}{c}  \eps \hat{F}_x  \\  \hat{F}_y  \end{array}\right)
,$$
was shown to vary strongly with $\Rey_\eps$ and not with $\eps$.
Our numerical experiments confirm these conclusions, see \Figurep{Dp--a} and~\ref{plot:draglift}.
We, however, observe much smaller values of the pressure differences $d_{\hat{p}}$ than those reported 
in~\cite{SzeriSnyder_2006},
as compared in~\Figurep{DpComparison}.
The~explanation for the discrepancy is not clear, 
as a~detailed discussion of the Navier--Stokes problem formulation and results is lacking%
\footnote{We suspect that boundary conditions 
different from~\eqref{eq:bc-artificial}, \eqref{nondim:bc-artificial}
could have been set on $\Gamma_\mathrm{in}$ and $\Gamma_\mathrm{out}$
in~\cite{SzeriSnyder_2006},
which might have caused cross flow and pressure gradients in the vicinity 
of both the artificial boundaries.
} in~\cite{SzeriSnyder_2006}.
Both the computed traction along the slider surface presented in \Figurep{draglift}
and the resulting lift for various slopes $h_2/h_1$ presented in Table~\ref{table:lift} 
show surprisingly small variation with $\eps$. 
It is worth noting that the values of $\hat{F}_y$ for $\eps=0.001$ in Table~\ref{table:lift}
coincide within the presented accuracy 
with the results obtained from the Reynolds equation, cf.~Table~1\ in~\cite{SzeriSnyder_2006},
while they provide a~surprisingly good approximation even to the problems with~$\eps=0.1$.

%
%


\subsection{Pressure-thickening, $\alpha^*>0$. Inappropriateness of the viscosity cut-off procedure and 
computational difficulties}
\label{sec:cutoff}
\begin{figure*}
\centering
\subfloat[Vertical component, $-\widetilde{\tens{T}}\hat{\vect{n}}\cdot\vect{e}_y$]{%
\begin{minipage}{0.45\textwidth}
	\begin{tikzpicture}
	\vspace{1em}%
		\begin{axis}[
			xlabel={ $\hat{x}$ },  
			ylabel={ $-\widetilde{\tens{T}}\hat{\vect{n}}\cdot\vect{e}_y$ },
			width={ 0.95\textwidth }, height={ 0.4\textheight },
			every axis y label/.style=
				{at={(ticklabel cs:0.7)},rotate=90,anchor=near ticklabel},
			unbounded coords=discard,
			font=\small,
			axis background/.style={fill=green!20},	
			legend pos=outer north east,
			cycle list name=linestyles*, 
		]
		\addplot+ table[x index=0,y index=3, smooth, solid, mark=none,col sep=space,trim cells=true] {results/1508b-cutoff/02013m6.Tnxy-slider.0001};
		  \addlegendentry{ no cut-off }
		\addplot+ table[x index=0,y index=3, smooth, solid, mark=none,col sep=space,trim cells=true] {results/1508b-cutoff/02233m6.Tnxy-slider.0001};
		  \addlegendentry{ $\bar{p}=2.2$ }
		\addplot+ table[x index=0,y index=3, smooth, solid, mark=none,col sep=space,trim cells=true] {results/1508b-cutoff/02133m6.Tnxy-slider.0001};
		  \addlegendentry{ $\bar{p}=2.0$ }
		\legend{}
		\end{axis}
	\end{tikzpicture}
\end{minipage}}
\hspace{0.05\textwidth}
\subfloat[Horizontal component, $-\widetilde{\tens{T}}\hat{\vect{n}}\cdot\vect{e}_x$]{%
\begin{minipage}{0.45\textwidth}
	\centering
	\begin{tikzpicture}
	\vspace{1em}%
		\begin{axis}[
			xlabel={ $\hat{x}$ },  
			ylabel={ $-\widetilde{\tens{T}}\hat{\vect{n}}\cdot\vect{e}_x$ },
			width={ 0.95\textwidth }, height={ 0.4\textheight },
			every axis y label/.style=
				{at={(ticklabel cs:0.7)},rotate=90,anchor=near ticklabel},
			unbounded coords=discard,
			font=\small,
			axis background/.style={fill=green!20},	
			legend pos=south east,
			cycle list name=linestyles*, 
		]
		\addplot+ table[x index=0,y index=2, smooth, solid, mark=none,col sep=space,trim cells=true] {results/1508b-cutoff/02013m6.Tnxy-slider.0001};
		  \addlegendentry{ no cut-off }
		\addplot+ table[x index=0,y index=2, smooth, solid, mark=none,col sep=space,trim cells=true] {results/1508b-cutoff/02233m6.Tnxy-slider.0001};
		  \addlegendentry{ $\bar{p}=2.2$ }
		\addplot+ table[x index=0,y index=2, smooth, solid, mark=none,col sep=space,trim cells=true] {results/1508b-cutoff/02133m6.Tnxy-slider.0001};
		  \addlegendentry{ $\bar{p}=2.0$ }
		\end{axis}
	\end{tikzpicture}
\end{minipage}}
\caption{Dimensionless traction along 
	$\hat{\Gamma}_{\mathrm{slider}}$, 
	for $\alpha^*=1.74$ 
	and different cut-off parameters
	($G^*=0$, $\eps=0.005$, $\Rey_\eps=0$)} 
\label{plot:A174}
\end{figure*}
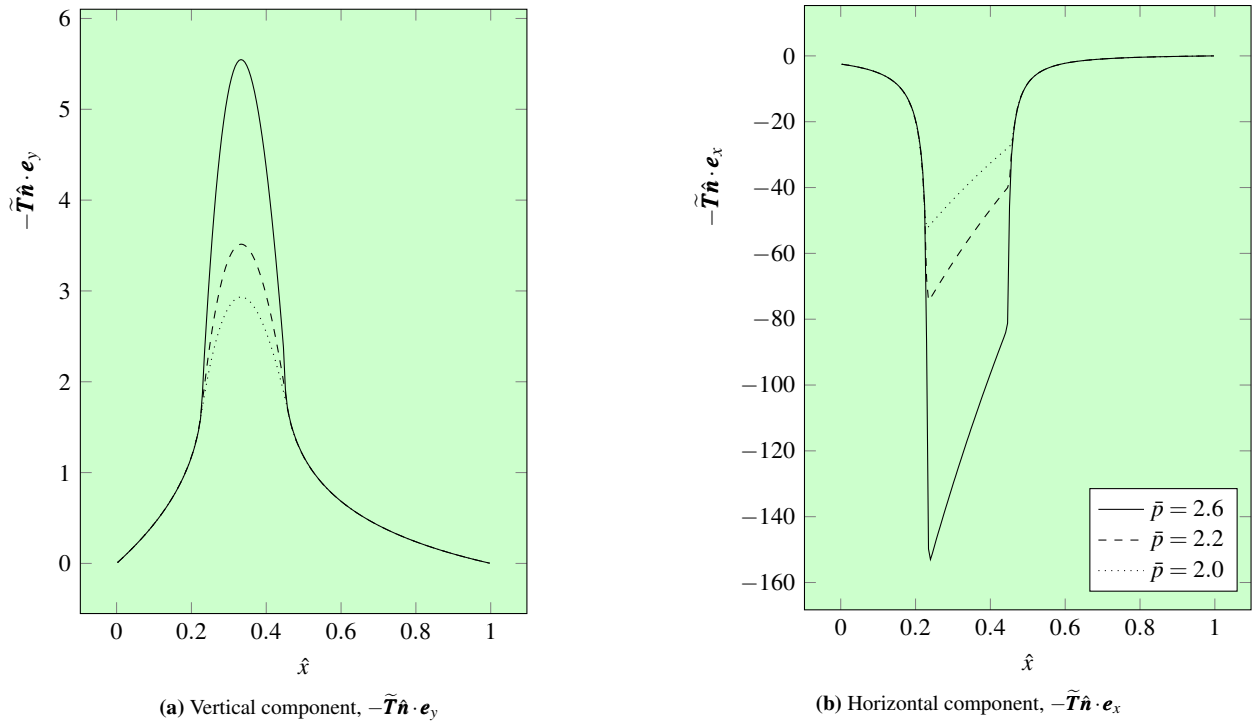
\begin{figure*}
\centering
\subfloat[Vertical component, $-\widetilde{\tens{T}}\hat{\vect{n}}\cdot\vect{e}_y$]{%
\begin{minipage}{0.45\textwidth}
	\begin{tikzpicture}
	\vspace{1em}%
		\begin{axis}[
			xlabel={ $\hat{x}$ },  
			ylabel={ $-\widetilde{\tens{T}}\hat{\vect{n}}\cdot\vect{e}_y$ },
			width={ 0.95\textwidth }, height={ 0.4\textheight },
			every axis y label/.style=
				{at={(ticklabel cs:0.7)},rotate=90,anchor=near ticklabel},
			unbounded coords=discard,
			font=\small,
			axis background/.style={fill=green!20},	
			legend pos=outer north east,
			cycle list name=linestyles*, 
		]
		\addplot+ table[x index=0,y index=3, smooth, solid, mark=none,col sep=space,trim cells=true] {results/1508b-cutoff/02425m6.Tnxy-slider.0001};
		  \addlegendentry{ $\bar{p}=2.6$ }
		\addplot+ table[x index=0,y index=3, smooth, solid, mark=none,col sep=space,trim cells=true] {results/1508b-cutoff/02216m6.Tnxy-slider.0001};
		  \addlegendentry{ $\bar{p}=2.2$ }
		\addplot+ table[x index=0,y index=3, smooth, solid, mark=none,col sep=space,trim cells=true] {results/1508b-cutoff/02116m6.Tnxy-slider.0001};
		  \addlegendentry{ $\bar{p}=2.0$ }
		\legend{}
		\end{axis}
	\end{tikzpicture}
\end{minipage}}
\hspace{0.05\textwidth}
\subfloat[Horizontal component, $-\widetilde{\tens{T}}\hat{\vect{n}}\cdot\vect{e}_x$]{%
\begin{minipage}{0.45\textwidth}
	\centering
	\begin{tikzpicture}
	\vspace{1em}%
		\begin{axis}[
			xlabel={ $\hat{x}$ },  
			ylabel={ $-\widetilde{\tens{T}}\hat{\vect{n}}\cdot\vect{e}_x$ },
			width={ 0.95\textwidth }, height={ 0.4\textheight },
			every axis y label/.style=
				{at={(ticklabel cs:0.7)},rotate=90,anchor=near ticklabel},
			unbounded coords=discard,
			font=\small,
			axis background/.style={fill=green!20},	
			legend pos=south east,
			cycle list name=linestyles*, 
		]
		\addplot+ table[x index=0,y index=2, smooth, solid, mark=none,col sep=space,trim cells=true] {results/1508b-cutoff/02425m6.Tnxy-slider.0001};
		  \addlegendentry{ $\bar{p}=2.6$ }
		\addplot+ table[x index=0,y index=2, smooth, solid, mark=none,col sep=space,trim cells=true] {results/1508b-cutoff/02216m6.Tnxy-slider.0001};
		  \addlegendentry{ $\bar{p}=2.2$ }
		\addplot+ table[x index=0,y index=2, smooth, solid, mark=none,col sep=space,trim cells=true] {results/1508b-cutoff/02116m6.Tnxy-slider.0001};
		  \addlegendentry{ $\bar{p}=2.0$ }
		\end{axis}
	\end{tikzpicture}
\end{minipage}}
\caption{Dimensionless traction along 
	$\hat{\Gamma}_{\mathrm{slider}}$, 
	for $\alpha^*=1.85$ 
	and three cut-off parameters
	(no solution without cut-off available)}
\label{plot:A185}
\end{figure*}
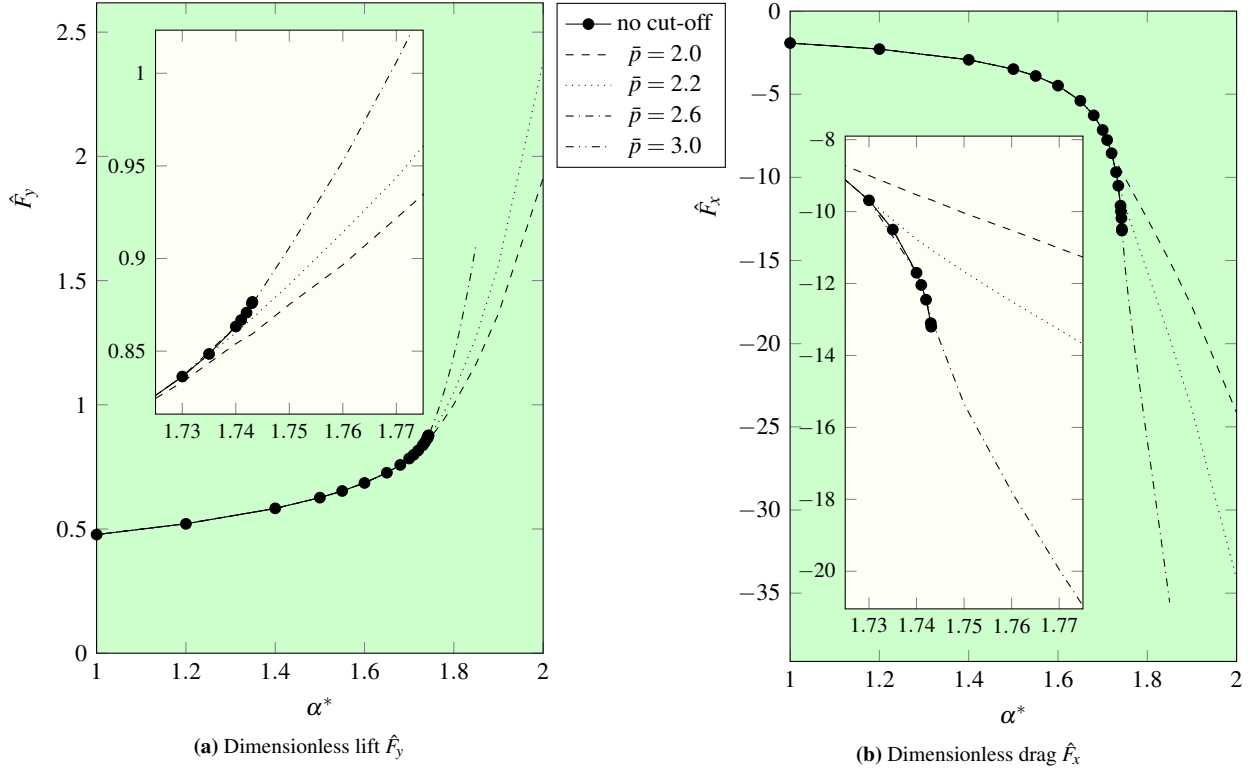
\begin{figure*}
\centering
\subfloat[Dimensionless lift $\hat{F}_y$]{%
\begin{minipage}{0.45\textwidth}
	\begin{tikzpicture}
	\vspace{1em}%
		\begin{axis}[
			xlabel={ $\alpha^*$ },  
			ylabel={ $\hat{F}_y$ },
			width={ 0.95\textwidth }, height={ 1.3\textwidth },
			every axis y label/.style=
				{at={(ticklabel cs:0.7)},rotate=90,anchor=near ticklabel},
			unbounded coords=discard,
			xmin=1.0, xmax=2.0, ymin=0.0,
			font=\small,
			axis background/.style={fill=green!20},	
			cycle list name=linestyles*, 
			legend pos=outer north east,
		]
		\addplot+ [mark=*] table[x index=0,y index=5, smooth, solid, col sep=space,trim cells=true] {results/1508b-cutoff/maxp0.0-mg6.txt};
		  \addlegendentry{ no cut-off }
		\addplot+ table[x index=0,y index=5, smooth, solid, col sep=space,trim cells=true] {results/1508b-cutoff/maxp2.0-mg6.txt};
		  \addlegendentry{ $\bar{p}=2.0$ }
		\addplot+ table[x index=0,y index=5, smooth, solid, col sep=space,trim cells=true] {results/1508b-cutoff/maxp2.2-mg6.txt};
		  \addlegendentry{ $\bar{p}=2.2$ }
		\addplot+ table[x index=0,y index=5, smooth, solid, col sep=space,trim cells=true] {results/1508b-cutoff/maxp2.6-mg6.txt};
		  \addlegendentry{ $\bar{p}=2.6$ }
		\addplot+ table[x index=0,y index=5, smooth, solid, col sep=space,trim cells=true] {results/1508b-cutoff/maxp3.0-mg6.txt};
		  \addlegendentry{ $\bar{p}=3.0$ }
		\end{axis}
	\node [align=right, anchor=west] at (0.65,5.5) {%
		\begin{tikzpicture}[baseline,trim axis left, trim axis right]
		\vspace{1em}%
			\begin{axis}[
				width={ 0.65\textwidth }, height={ 0.85\textwidth },
				unbounded coords=discard,
				xmin=1.725, xmax=1.775,
				font=\footnotesize,
				axis background/.style={fill=yellow!5},	
				xtick={1.73,1.74,1.75,1.76,1.77},
				cycle list name=linestyles*, 
			]
			\addplot+ [mark=*] table[x index=0,y index=5, smooth, solid, col sep=space,trim cells=true] {results/1508b-cutoff/maxp0.0-mg6.txt};
			\addplot+ table[x index=0,y index=5, smooth, solid, col sep=space,trim cells=true] {results/1508b-cutoff/maxp2.0-mg6.txt};
			\addplot+ table[x index=0,y index=5, smooth, solid, col sep=space,trim cells=true] {results/1508b-cutoff/maxp2.2-mg6.txt};
			\addplot+ table[x index=0,y index=5, smooth, solid, col sep=space,trim cells=true] {results/1508b-cutoff/maxp2.6-mg6.txt};
			\addplot+ table[x index=0,y index=5, smooth, solid, col sep=space,trim cells=true] {results/1508b-cutoff/maxp3.0-mg6.txt};
			\end{axis}
		\end{tikzpicture}
	};
	\end{tikzpicture}
\end{minipage}}
\hspace{0.05\textwidth}
\subfloat[Dimensionless drag $\hat{F}_x$]{%
\begin{minipage}{0.45\textwidth}
	\centering
	\begin{tikzpicture}
	\vspace{1em}%
		\begin{axis}[
			xlabel={ $\alpha^*$ },  
			ylabel={ $\hat{F}_x$ },
			width={ 0.95\textwidth }, height={ 1.3\textwidth },
			every axis y label/.style=
				{at={(ticklabel cs:0.7)},rotate=90,anchor=near ticklabel},
			unbounded coords=discard,
			xmin=1.0, xmax=2.0, ymax=0.0,
			font=\small,
			axis background/.style={fill=green!20},	
			legend pos=outer north east,
			cycle list name=linestyles*, 
		]
		\addplot+ [mark=*] table[x index=0,y index=6, smooth, solid, col sep=space,trim cells=true] {results/1508b-cutoff/maxp0.0-mg6.txt};
		  \addlegendentry{ no cut-off }
		\addplot+ table[x index=0,y index=6, smooth, solid, col sep=space,trim cells=true] {results/1508b-cutoff/maxp2.0-mg6.txt};
		  \addlegendentry{ $\bar{p}=2.0$ }
		\addplot+ table[x index=0,y index=6, smooth, solid, col sep=space,trim cells=true] {results/1508b-cutoff/maxp2.2-mg6.txt};
		  \addlegendentry{ $\bar{p}=2.2$ }
		\addplot+ table[x index=0,y index=6, smooth, solid, col sep=space,trim cells=true] {results/1508b-cutoff/maxp2.6-mg6.txt};
		  \addlegendentry{ $\bar{p}=2.6$ }
		\addplot+ table[x index=0,y index=6, smooth, solid, col sep=space,trim cells=true] {results/1508b-cutoff/maxp3.0-mg6.txt};
		  \addlegendentry{ $\bar{p}=3.0$ }
		\legend{}
		\end{axis}
	\node [align=right, anchor=west] at (0.6,3.7) {%
		\begin{tikzpicture}[baseline,trim axis left, trim axis right]
		\vspace{1em}%
			\begin{axis}[
				width={ 0.6\textwidth }, height={ 1\textwidth },
				unbounded coords=discard,
				xmin=1.725, xmax=1.775,
				font=\footnotesize,
				axis background/.style={fill=yellow!5},	
				xtick={1.73,1.74,1.75,1.76,1.77},
				cycle list name=linestyles*, 
			]
			\addplot+ [mark=*] table[x index=0,y index=6, smooth, solid, col sep=space,trim cells=true] {results/1508b-cutoff/maxp0.0-mg6.txt};
			\addplot+ table[x index=0,y index=6, smooth, solid, col sep=space,trim cells=true] {results/1508b-cutoff/maxp2.0-mg6.txt};
			\addplot+ table[x index=0,y index=6, smooth, solid, col sep=space,trim cells=true] {results/1508b-cutoff/maxp2.2-mg6.txt};
			\addplot+ table[x index=0,y index=6, smooth, solid, col sep=space,trim cells=true] {results/1508b-cutoff/maxp2.6-mg6.txt};
			\addplot+ table[x index=0,y index=6, smooth, solid, col sep=space,trim cells=true] {results/1508b-cutoff/maxp3.0-mg6.txt};
			\end{axis}
		\end{tikzpicture}
	};
	\end{tikzpicture}
\end{minipage}}
\caption{Dimensionless force acting on the slider, 
	for different cut-off parameters 
	($\Rey_\eps=0$, $G^*=0$, $h_2/h_1=2$, $\eps=0.005$)}
\label{plot:fyfx-cutoff}
\end{figure*}
\begin{figure*}[h!]
\centering
\subfloat[$\hat{\eta}$ (unaltered)]{
	\includegraphics[width=0.45\textwidth,trim=200 300 0 250,clip=true]{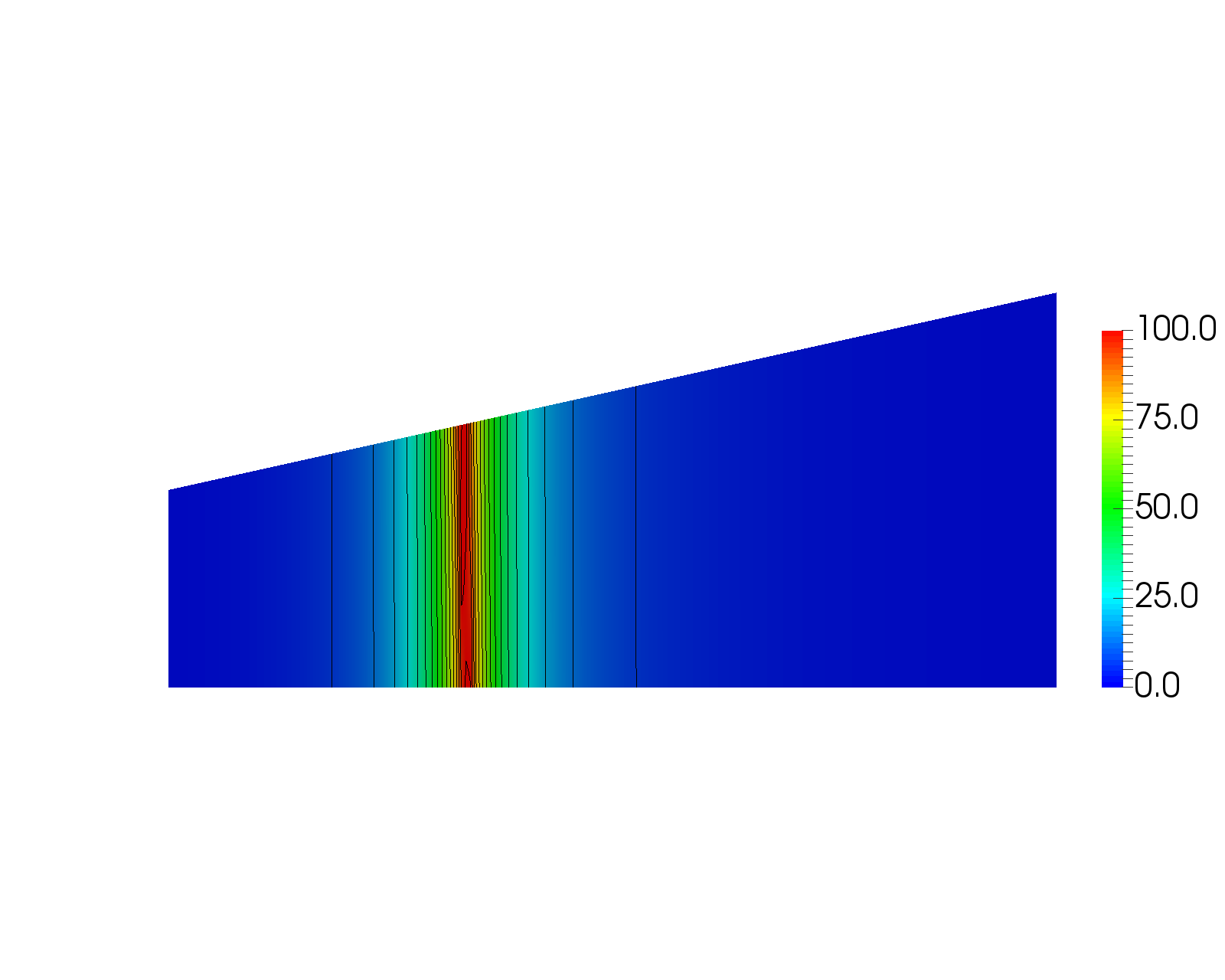}
}
\hspace{2ex}
\subfloat[$\hat{\eta}$ ($\bar{p}=2.0$)]{
	\includegraphics[width=0.45\textwidth,trim=200 300 0 250,clip=true]{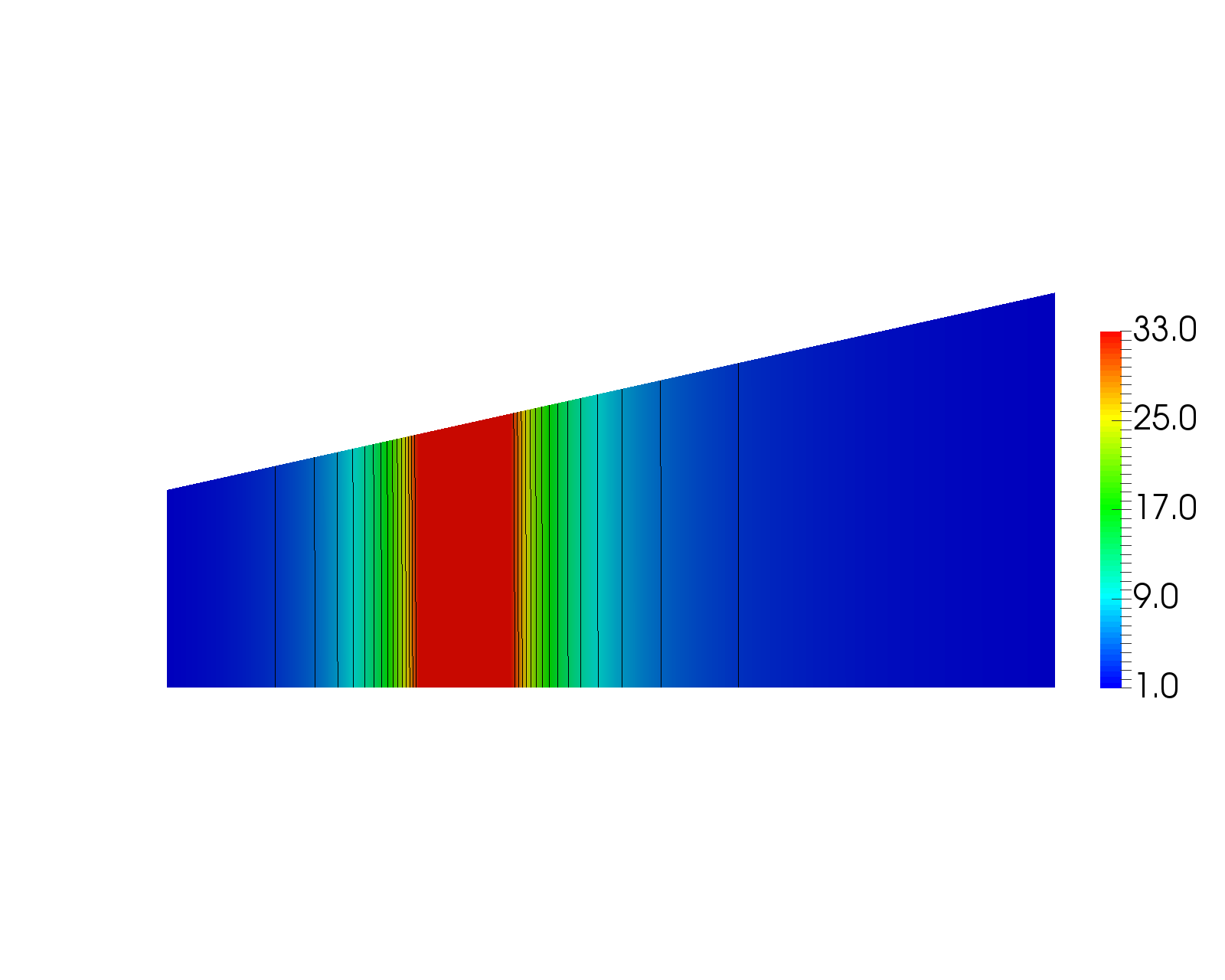}
}
\\ 
\subfloat[$\hat{p}$ (unaltered)]{
	\includegraphics[width=0.45\textwidth,trim=200 300 0 250,clip=true]{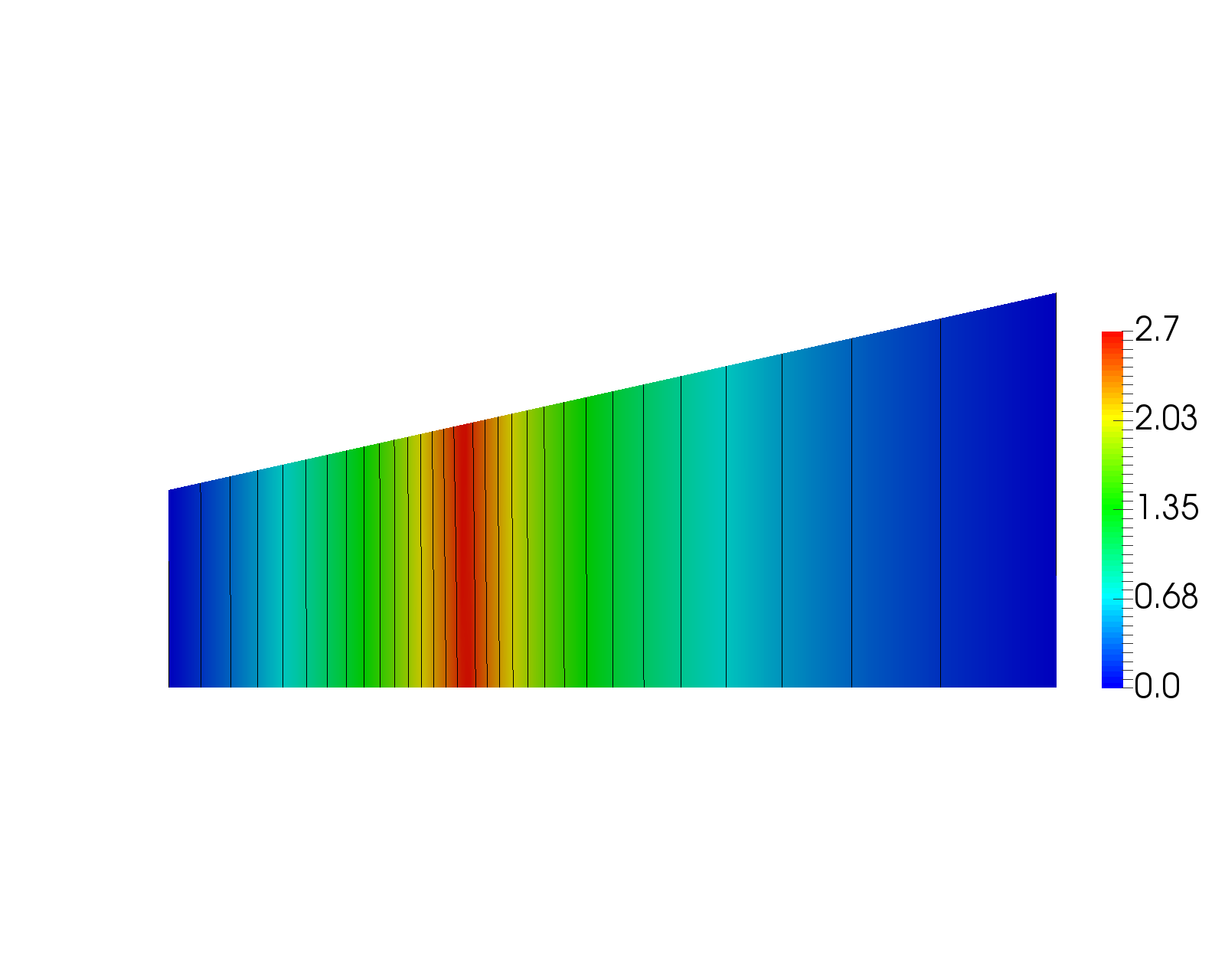}
}
\hspace{2ex}
\subfloat[$\hat{p}$ ($\bar{p}=2.0$)]{
	\includegraphics[width=0.45\textwidth,trim=200 300 0 250,clip=true]{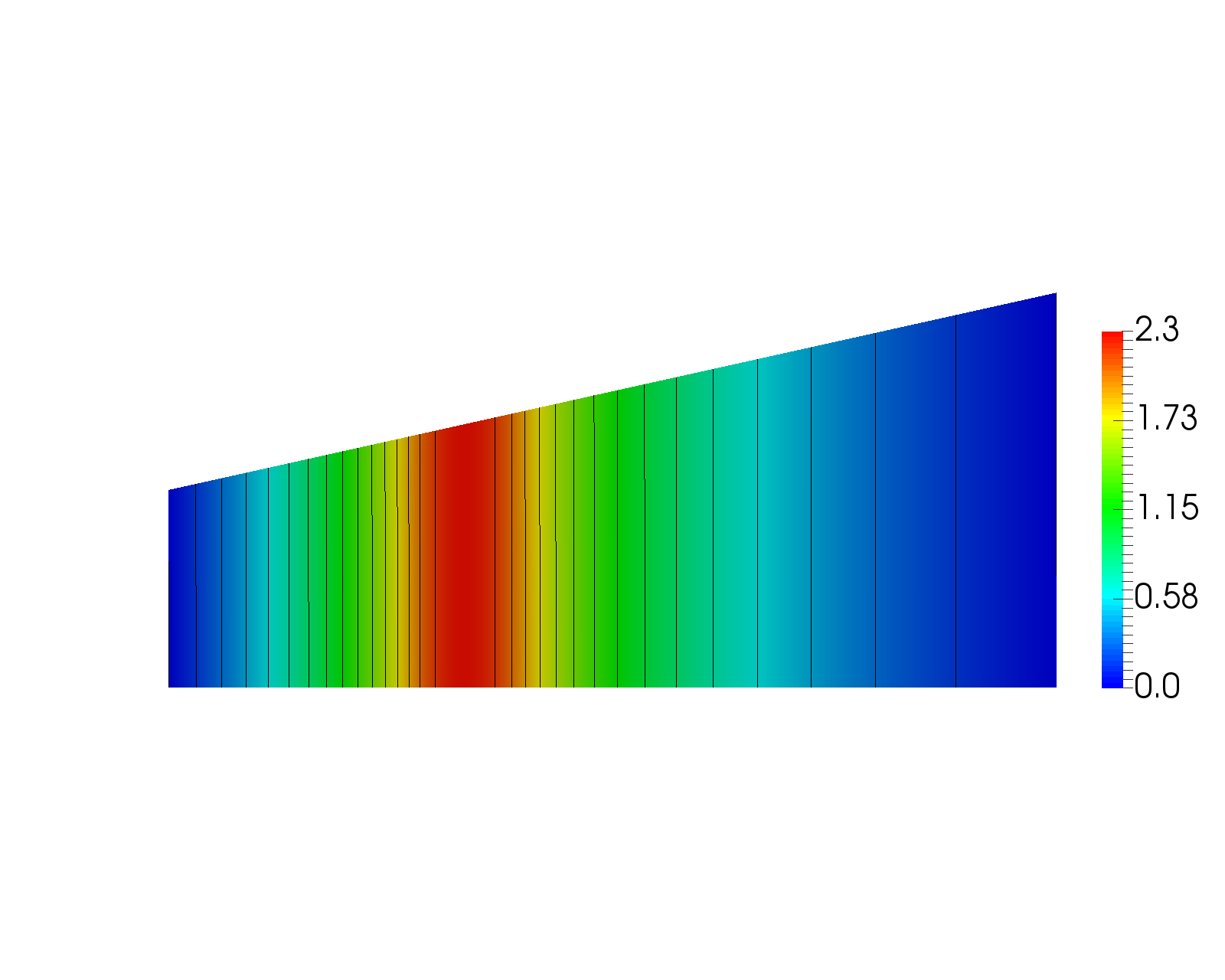}
}
\\ 
\subfloat[$\hat{u}$ (unaltered)]{
	\includegraphics[width=0.45\textwidth,trim=200 300 0 250,clip=true]{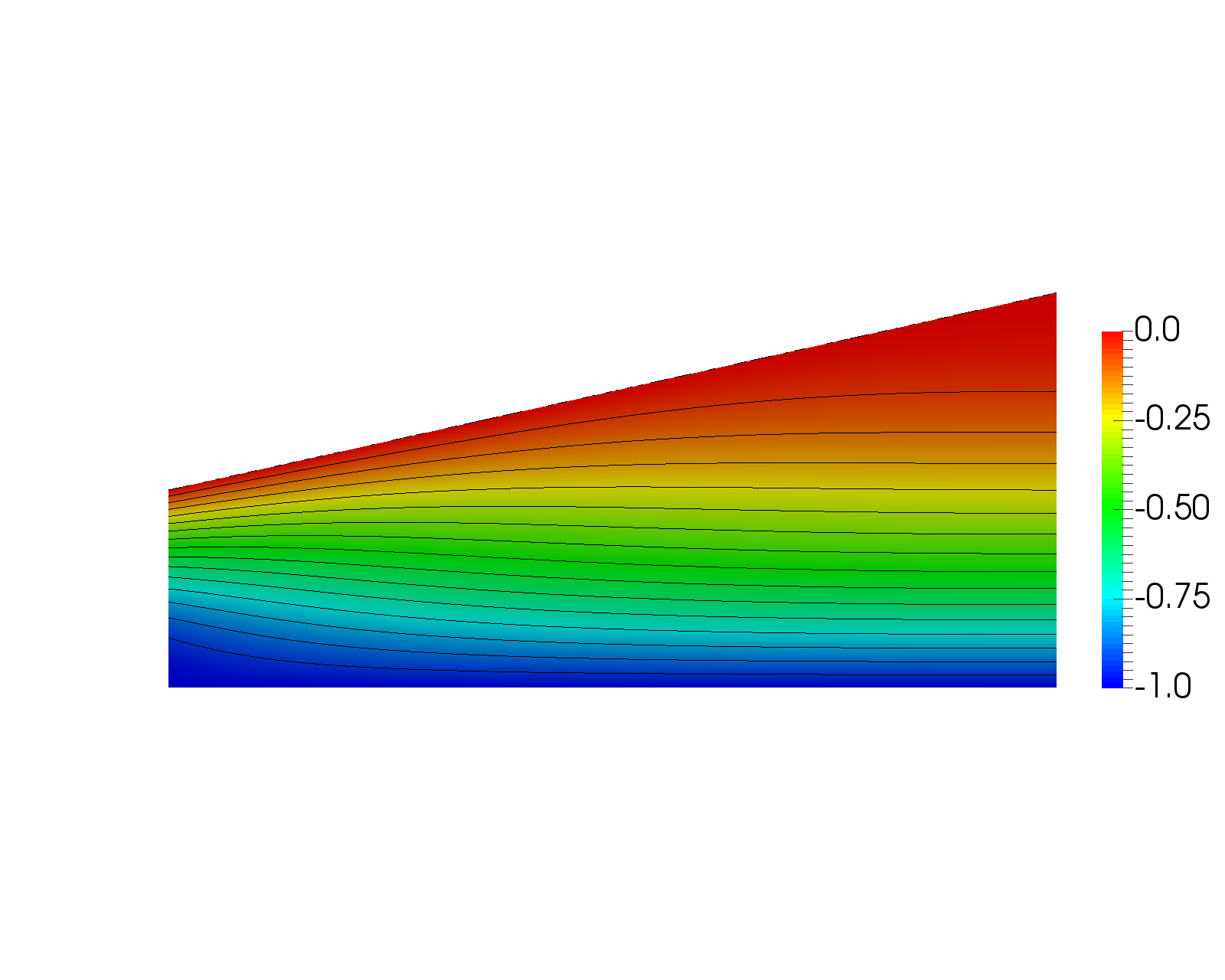}
}
\hspace{2ex}
\subfloat[$\hat{u}$ ($\bar{p}=2.0$)]{
	\includegraphics[width=0.45\textwidth,trim=200 300 0 250,clip=true]{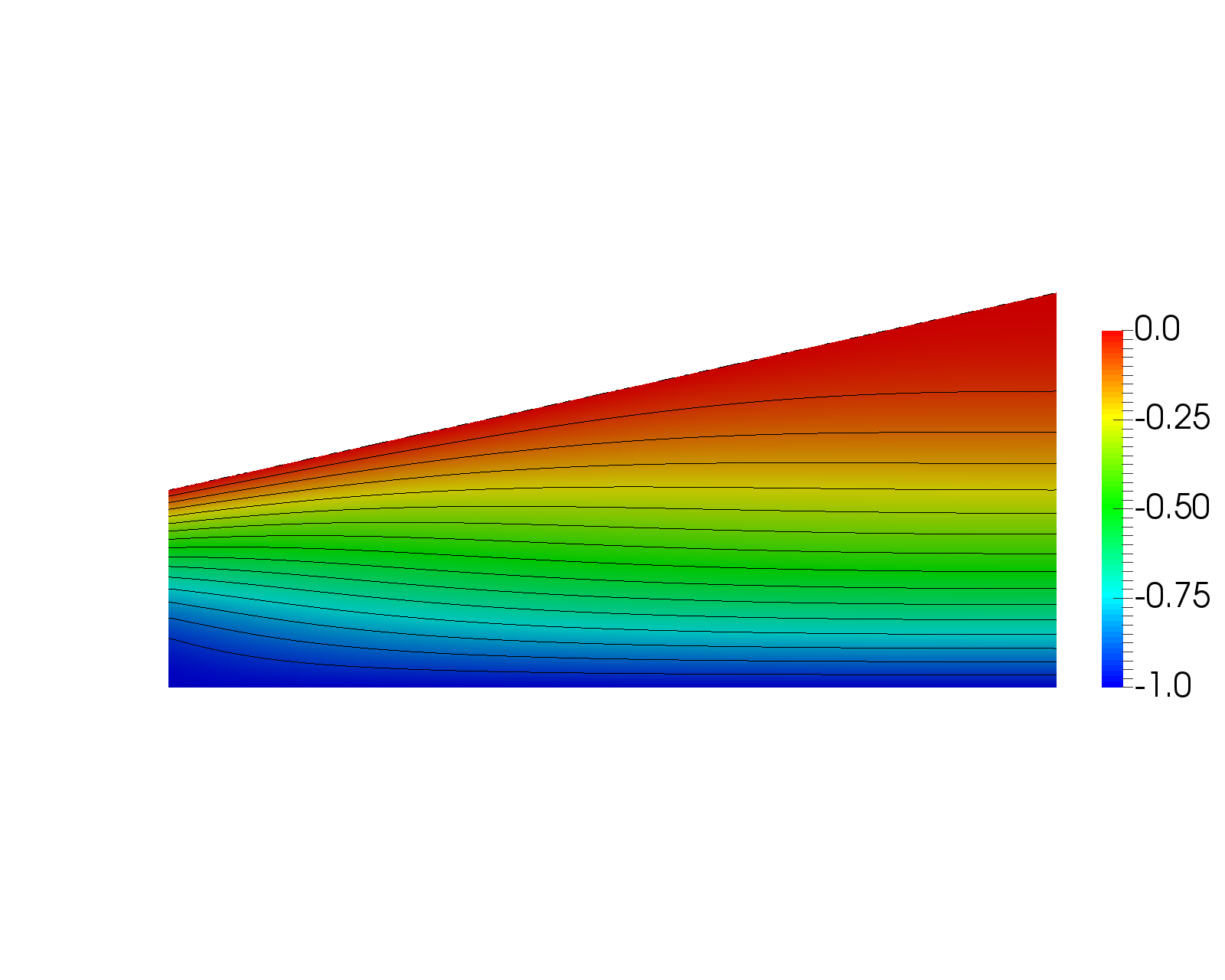}
}
\\ 
\subfloat[$\hat{v}$ (unaltered)]{
	\includegraphics[width=0.45\textwidth,trim=200 300 0 250,clip=true]{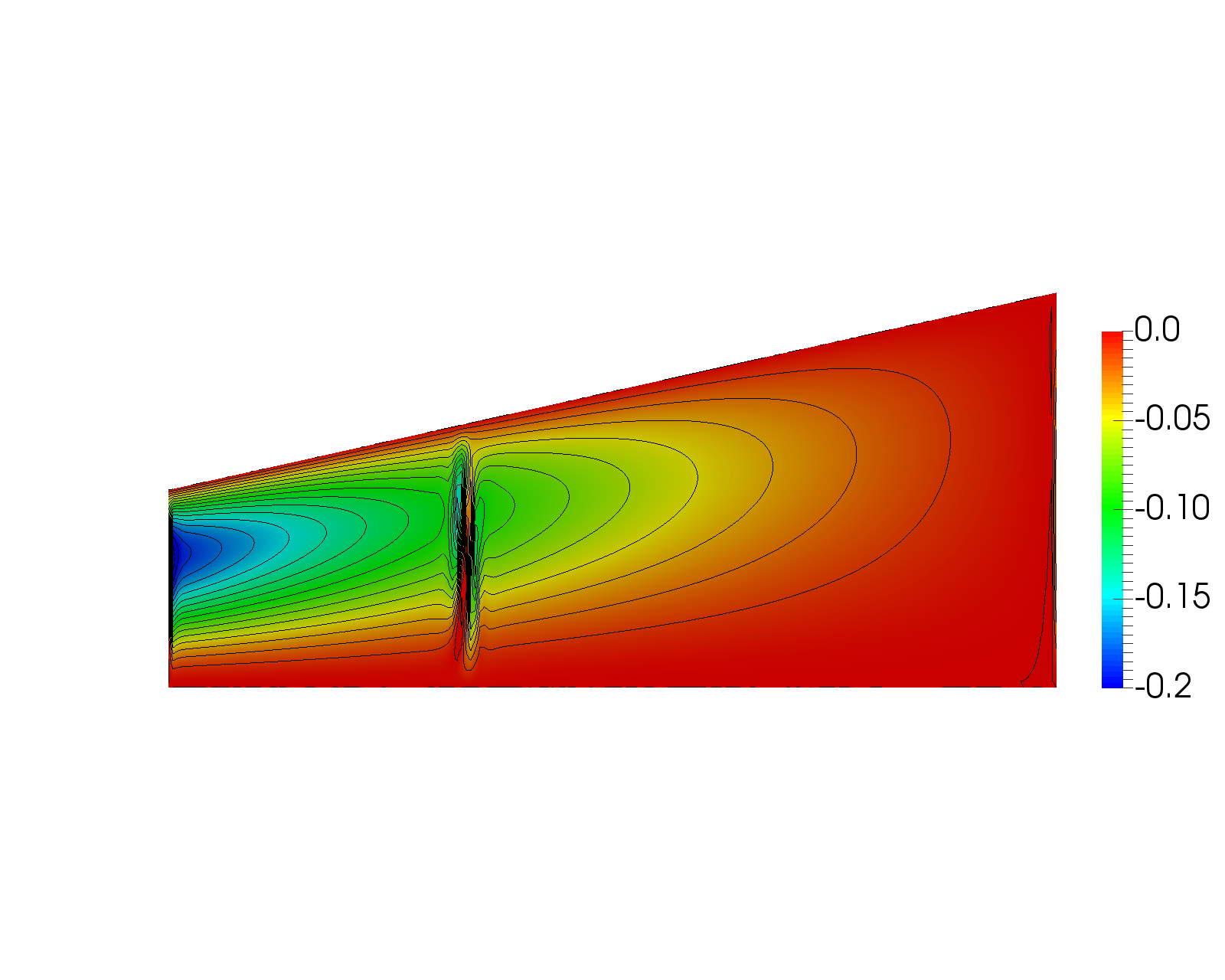}
}
\hspace{2ex}
\subfloat[$\hat{v}$ ($\bar{p}=2.0$)]{
	\includegraphics[width=0.45\textwidth,trim=200 300 0 250,clip=true]{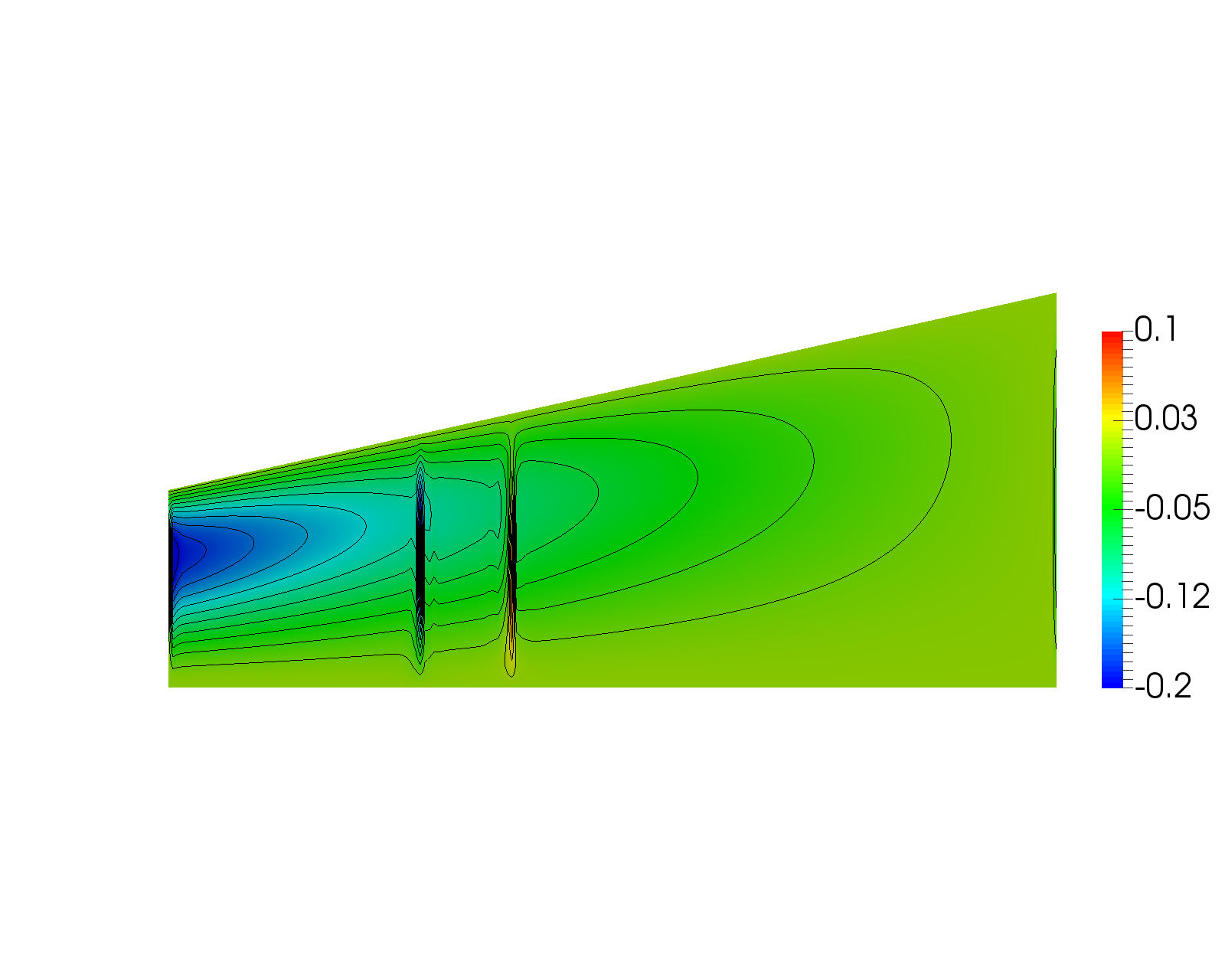}
}
\caption{Dimensionless viscosity $\hat\eta$, pressure $\hat{p}$ and velocity $\hat{\vect{u}}$
	for $\alpha^*=1.743$,
	for unaltered viscosity (left) and using the cut off (right),
	($G^*=0.0$, $\eps=0.005$, $\Rey_\eps=0$)}
\label{fig:comparison-A174-cutoff}
\end{figure*}
In all the remaining examples, we take $h_2/h_1=2$.
When $\alpha^*>0$, the fluid is pressure-thickening.
For clarity of exposition, we start with $G^*=0$, reducing \eqref{nondim:expCY} to 
the exponential pressure--viscosity model~$\hat{\eta}=e^{\alpha^*\hat{p}}$.
As $\alpha^*$ increases, the other parameters being fixed, the pressure peak generated
within the plane slider flow grows; the non-linear character of the system is emphasized
and the discrete problem is more difficult to handle. 
Eventually, for $\alpha^*$ large enough, \eqref{eq:dSdp} is violated, bringing about a~failure of the numerical scheme.
This observation seems in accordance with what has been encountered by other researchers,
cf.~\cite{Almqvist2002,Knauf2013}.

In order to prevent the failure of the numerical calculation, 
a~technique practiced by some researchers has been to cut off the viscosity
by employing, e.g.,
\begin{equation}\label{eq:cutoff}
	\hat{\eta}_{\bar{p}} = \hat{\eta}( \min\{ \hat{p}, \bar{p} \}, \tr\hat{\tens{D}}^2 )
\end{equation}
instead of $\hat{\eta}$,
or similarly by imposing a~restriction on the norm of stress by employing, e.g., 
$$	\hat{\eta}_{\bar{S}} 
	= \min\{ \hat{\eta}(\cdot,\cdot) \tr\hat{\tens{D}}^2 , \bar{S} \}
	  /\tr\hat{\tens{D}}^2
.$$
To pick some examples wherein such a~cut off has been appealed to, we refer to~\cite{%
Almqvist2002,%
Knauf2013,%
LiDaviesPhillips2000,%
DaviesLi1994,%
GwynllywDaviesPhillips1996%
}. 
Doing so, one can ensure $\partial\tens{S}/\partial p$ to remain bounded and,
by choosing suitable threshold parameter $\bar{p}$ or $\bar{S}$, to keep \eqref{eq:dSdp} fulfilled at least for bounded shear rates.
In particular, using \eqref{eq:cutoff} and considering for instance \eqref{nondim:expCY}
and given $\bar{D}$ and $C$, one can find $\bar{p}$ such that \eqref{eq:dSdp} holds for any $\tr\tens{D}^2\leq \bar{D}$.
One should notice, however, examining \eqref{nondim:expCY} with any $\alpha^*>0$ and $G^*,\;\beta^*\geq0$,
that for any choice of $\bar{p}$, \eqref{eq:dSdp} is still violated for sufficiently large shear rates.

Surprisingly, the possible sensitivity of the solution and of the derived quantities of interest
on the cut-off parameter has not been investigated in the literature so far,
to the best of our knowledge.
We provide the following set of numerical experiments to document 
that, once the cut-off takes effect, the results depend sorely on the parameter $\bar{p}$.

For convenience, the comparison is presented for $\eps=0.005$ and $\Rey_\eps=0$,
but we observed that the behaviour is qualitatively the same for higher values of these parameters as well.
The dimensionless traction along the slider surface is presented in \Figurep{A174}, 
where the results for the unaltered viscosity and for two different cut-off parameters
are compared for $\alpha^*=1.74$.
Note that while the vertical component 
(which corresponds almost exactly to the pressure distribution and sums up to the resulting lift force)
does not vary considerably in this example, 
the horizontal component (which determines the resulting friction)
is affected substantially.
The differences are even more pronounced in \Figurep{A185}. 
For $\alpha^*=1.85$, we were unable to find any solution with unaltered viscosity, 
the condition \eqref{eq:dSdp} being eventually violated while attempting to solve the discrete nonlinear system.
Therefore we only present the results for three values of $\bar{p}$,
showing a~marked variation in both components of the traction.

In terms of the resulting force as a~function of $\alpha^*$,
the comparison is presented in \Figurep{fyfx-cutoff}.
For $\alpha^*<1.72$, the maximum of the resulting dimensionless pressure does not reach 
the lowest cut-off threshold $\bar{p}=2.0$, 
hence all the curves plotted in \Figurep{fyfx-cutoff} 
coincide up to that value. 
With the unaltered model, we were only able to proceed up to $\alpha^*=1.743$,
same as in the case with $\bar{p}=3.0$. 
With $\bar{p}=2.6$, the computation fails for $\alpha^*\geq 1.86$.
\Figurep{fyfx-cutoff} illustrates that once the viscosity cut-off takes effect, 
the resulting force is altered significantly.
We may conclude, that while the lower cut-off parameters may seem to add to the 
robustness of the computation, they actually entail strikingly different results 
depending on the choice of $\bar{p}$, making such solutions unreliable.

To enhance the illustration, we present the comparison of the solutions
with unaltered viscosity and with the cut-off defined by $\bar{p}=2.0$
in \Figure{comparison-A174-cutoff},
for the case $\alpha^*=1.743$.
The dimensionless pressure peak in the unaltered case reaches $\hat{p}=2.7$ (c),
while with the cut viscosity it is lowered to $\hat{p}=2.3$ (d).
The difference is more pronounced in the corresponding maximal values of the viscosity,
the peak value $\hat{\eta}=100$ (a) is lowered to $\hat{\eta}=33$ (b),
the viscosity now being constant in a~substantial part of the domain around the pressure peak.
Moreover, while the differences in the horizontal component $\hat{u}$ of the dimensionless velocity
can not be distinguished visually (e,f), there is a~striking difference in its vertical component $\hat{v}$.
While the unaltered case (g) results in a~distinct rib in $\hat{v}$,
related to the non-negligible variation of the viscosity with pressure 
in the vicinity of the pressure and viscosity peak,
in the altered case (h) the rib is replaced by two stronger artefacts 
positioned where the artificial viscosity cut off takes effect.

In all what follows, we use the unaltered viscosity \eqref{nondim:expCY}.

\subsection{Pressure variations across the film induced by pressure-thickening}
\label{sec:dp}
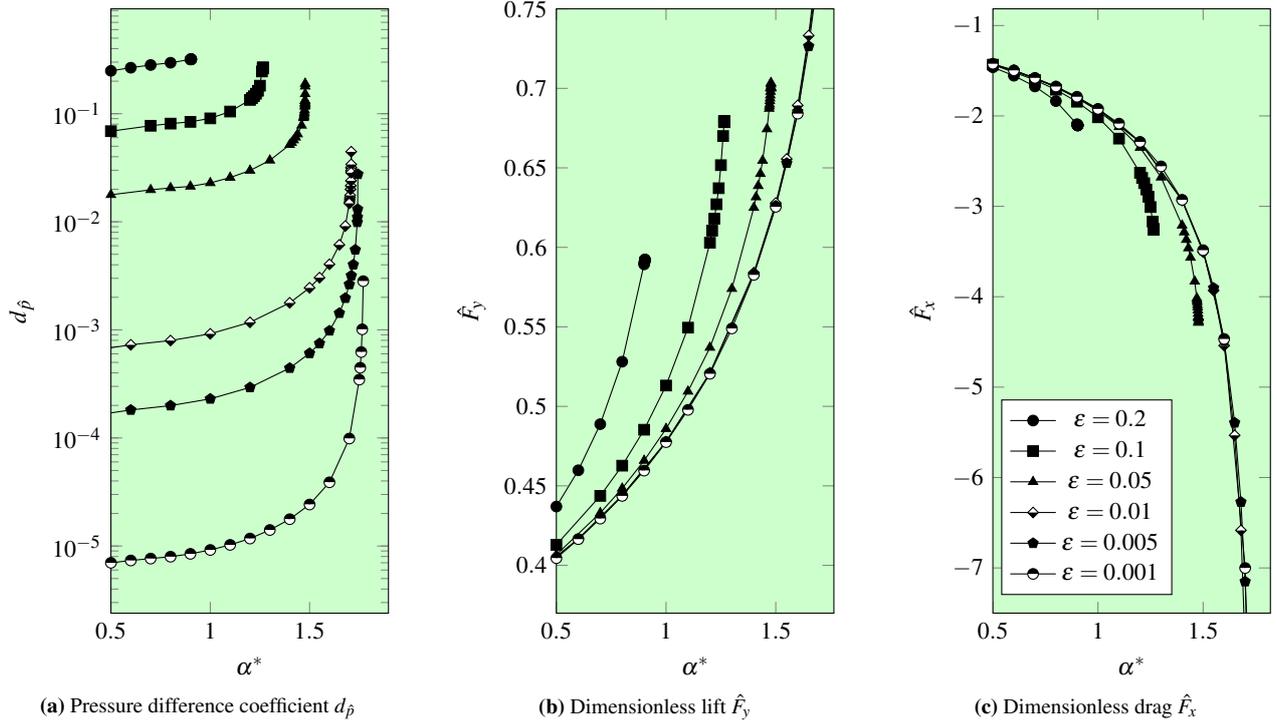
\begin{figure*}
\centering
\subfloat[Pressure difference coefficient $d_{\hat{p}}$]{%
\centering
	\begin{tikzpicture}
		\begin{semilogyaxis}[
			width={0.3\textwidth}, height={0.4\textheight},
			xlabel=$\alpha^*$,  
			ylabel=$d_{\hat{p}}$,
			unbounded coords=discard,
			xmin=0.5,
			font=\small,
			axis background/.style={fill=green!20},	
			legend style={anchor=south west,at={(0.1,0.05)},reverse legend}, 
			cycle list name=mark list* 
		]
		\addplot+ table[x index=3,y index=5, smooth, solid, mark=.] {results/1602a-dp-eps-alpha/dp-e2e-1-G0.0-Re0.0.txt};
			\addlegendentry{ $\eps=0.2$ };
		\addplot+ table[x index=3,y index=5, smooth, solid, mark=.] {results/1602a-dp-eps-alpha/dp-e1e-1-G0.0-Re0.0.txt};
			\addlegendentry{ $\eps=0.1$ };
		\addplot+ table[x index=3,y index=5, smooth, solid, mark=.] {results/1602a-dp-eps-alpha/dp-e5e-2-G0.0-Re0.0.txt};
			\addlegendentry{ $\eps=0.05$ };
		\addplot+ table[x index=3,y index=5, smooth, solid, mark=.] {results/1602a-dp-eps-alpha/dp-e1e-2-G0.0-Re0.0.txt};
			\addlegendentry{ $\eps=0.01$ };
		\addplot+ table[x index=3,y index=5, smooth, solid, mark=.] {results/1602a-dp-eps-alpha/dp-e5e-3-G0.0-Re0.0.txt};
			\addlegendentry{ $\eps=0.005$ };
		\addplot+ table[x index=3,y index=5, smooth, solid, mark=.] {results/1602a-dp-eps-alpha/dp-e1e-3-G0.0-Re0.0.txt};
			\addlegendentry{ $\eps=0.001$ };
		\legend{};
		\end{semilogyaxis}
	\end{tikzpicture}
}\hspace{0.04\textwidth}\subfloat[Dimensionless lift $\hat{F}_y$]{%
	\centering
	\begin{tikzpicture}
		\begin{axis}[
			width={0.3\textwidth}, height={0.4\textheight},
			xlabel=$\alpha^*$,  
			ylabel=$\hat{F}_y$,
			unbounded coords=discard,
			xmin=0.5,
			ymax=0.75,
			font=\small,
			axis background/.style={fill=green!20},	
			legend style={anchor=south west,at={(0.1,0.05)},reverse legend}, 
			cycle list name=mark list*, 
		]
		\addplot+ table[x index=3,y index=7, smooth, solid, mark=.] {results/1602a-dp-eps-alpha/dp-e2e-1-G0.0-Re0.0.txt};
			\addlegendentry{ $\eps=0.2$ };
		\addplot+ table[x index=3,y index=7, smooth, solid, mark=.] {results/1602a-dp-eps-alpha/dp-e1e-1-G0.0-Re0.0.txt};
			\addlegendentry{ $\eps=0.1$ };
		\addplot+ table[x index=3,y index=7, smooth, solid, mark=.] {results/1602a-dp-eps-alpha/dp-e5e-2-G0.0-Re0.0.txt};
			\addlegendentry{ $\eps=0.05$ };
		\addplot+ table[x index=3,y index=7, smooth, solid, mark=.] {results/1602a-dp-eps-alpha/dp-e1e-2-G0.0-Re0.0.txt};
			\addlegendentry{ $\eps=0.01$ };
		\addplot+ table[x index=3,y index=7, smooth, solid, mark=.] {results/1602a-dp-eps-alpha/dp-e5e-3-G0.0-Re0.0.txt};
			\addlegendentry{ $\eps=0.005$ };
		\addplot+ table[x index=3,y index=7, smooth, solid, mark=.] {results/1602a-dp-eps-alpha/dp-e1e-3-G0.0-Re0.0.txt};
			\addlegendentry{ $\eps=0.001$ };
		\legend{};
		\end{axis}
	\end{tikzpicture}
}\hspace{0.04\textwidth}\subfloat[Dimensionless drag $\hat{F}_x$]{
	\centering
	\begin{tikzpicture}
		\begin{axis}[
			width={0.3\textwidth}, height={0.4\textheight},
			xlabel=$\alpha^*$,  
			ylabel=$\hat{F}_x$,
			unbounded coords=discard,
			xmin=0.5,
			ymin=-7.5,
			font=\small,
			axis background/.style={fill=green!20},	
			legend pos=south west,
			cycle list name=mark list* 
		]
		\addplot+ table[x index=3,y index=8, smooth, solid, mark=.] {results/1602a-dp-eps-alpha/dp-e2e-1-G0.0-Re0.0.txt};
			\addlegendentry{ $\eps=0.2$ };
		\addplot+ table[x index=3,y index=8, smooth, solid, mark=.] {results/1602a-dp-eps-alpha/dp-e1e-1-G0.0-Re0.0.txt};
			\addlegendentry{ $\eps=0.1$ };
		\addplot+ table[x index=3,y index=8, smooth, solid, mark=.] {results/1602a-dp-eps-alpha/dp-e5e-2-G0.0-Re0.0.txt};
			\addlegendentry{ $\eps=0.05$ };
		\addplot+ table[x index=3,y index=8, smooth, solid, mark=.] {results/1602a-dp-eps-alpha/dp-e1e-2-G0.0-Re0.0.txt};
			\addlegendentry{ $\eps=0.01$ };
		\addplot+ table[x index=3,y index=8, smooth, solid, mark=.] {results/1602a-dp-eps-alpha/dp-e5e-3-G0.0-Re0.0.txt};
			\addlegendentry{ $\eps=0.005$ };
		\addplot+ table[x index=3,y index=8, smooth, solid, mark=.] {results/1602a-dp-eps-alpha/dp-e1e-3-G0.0-Re0.0.txt};
			\addlegendentry{ $\eps=0.001$ };
		\end{axis}
	\end{tikzpicture}
}
\caption{Dimensionless force $\hat{\vect{F}}$ and pressure difference coeficient $d_{\hat{p}}$, 
	variation with $\eps$ and $\alpha^*$
	($h_2/h_1=2$, $\Rey_\eps=0$, $G^*=0$)} 
\label{plot:dp-eps-alpha}
\end{figure*}
\begin{figure*}
\centering
\subfloat[Pressure difference coefficient $d_{\hat{p}}$]{%
\centering
	\begin{tikzpicture}
		\begin{semilogyaxis}[
			width={0.3\textwidth}, height={0.4\textheight},
			xlabel=$\alpha^*$,  
			ylabel=$d_{\hat{p}}$,
			unbounded coords=discard,
			xmin=1.0,
			font=\small,
			axis background/.style={fill=green!20},	
			legend pos=north west,
			cycle list name=mark list* 
		]
		\addplot+ table[x index=3,y index=5, smooth, solid, mark=.] {results/1602a-dp-eps-alpha/dp-e5e-3-G0.0-Re0.0.txt};
		\addplot+ table[x index=3,y index=5, smooth, solid, mark=.] {results/1602a-dp-eps-alpha/dp-e5e-3-G0.0-Re2.5.txt};
		\addplot+ [dotted] table[x index=3,y index=5, smooth, solid, mark=.] {results/1602a-dp-eps-alpha/dp-e5e-3-G0.0-Re5.0.txt};
		\end{semilogyaxis}
	\end{tikzpicture}
}\hspace{0.03\textwidth}\subfloat[Dimensionless lift $\hat{F}_y$]{%
	\centering
	\begin{tikzpicture}
		\begin{axis}[
			width={0.3\textwidth}, height={0.4\textheight},
			xlabel=$\alpha^*$,  
			ylabel=$\hat{F}_y$,
			unbounded coords=discard,
			xmin=1.0,
			font=\small,
			axis background/.style={fill=green!20},	
			legend pos=north west,
			cycle list name=mark list* 
		]
		\addplot+ table[x index=3,y index=7, smooth, solid, mark=.] {results/1602a-dp-eps-alpha/dp-e5e-3-G0.0-Re0.0.txt};
		\addplot+ table[x index=3,y index=7, smooth, solid, mark=.] {results/1602a-dp-eps-alpha/dp-e5e-3-G0.0-Re2.5.txt};
		\addplot+ [dotted] table[x index=3,y index=7, smooth, solid, mark=.] {results/1602a-dp-eps-alpha/dp-e5e-3-G0.0-Re5.0.txt};
		\end{axis}
	\end{tikzpicture}
}\hspace{0.03\textwidth}\subfloat[Dimensionless drag $\hat{F}_x$]{
	\centering
	\begin{tikzpicture}
		\begin{axis}[
			width={0.3\textwidth}, height={0.4\textheight},
			xlabel=$\alpha^*$,  
			ylabel=$\hat{F}_x$,
			unbounded coords=discard,
			xmin=1.0,
			font=\small,
			axis background/.style={fill=green!20},	
			legend pos=south west,
			cycle list name=mark list* 
		]
		\addplot+ table[x index=3,y index=8, smooth, solid, mark=.] {results/1602a-dp-eps-alpha/dp-e5e-3-G0.0-Re0.0.txt};
			\addlegendentry{ $\Rey_\eps=0$ };
		\addplot+ table[x index=3,y index=8, smooth, solid, mark=.] {results/1602a-dp-eps-alpha/dp-e5e-3-G0.0-Re2.5.txt};
			\addlegendentry{ $\Rey_\eps=2.5$ };
		\addplot+ [dotted] table[x index=3,y index=8, smooth, solid, mark=.] {results/1602a-dp-eps-alpha/dp-e5e-3-G0.0-Re5.0.txt};
			\addlegendentry{ $\Rey_\eps=5$ };
		\end{axis}
	\end{tikzpicture}
}
\caption{Dimensionless force $\hat{\vect{F}}$ and pressure difference coeficient $d_{\hat{p}}$, 
	for various $\Rey_\eps$ and $\alpha^*$
	($h_2/h_1=2$, $G^*=0$, $\eps=0.005$)} 
\label{plot:dp-eps-alpha-Rey}
\end{figure*}
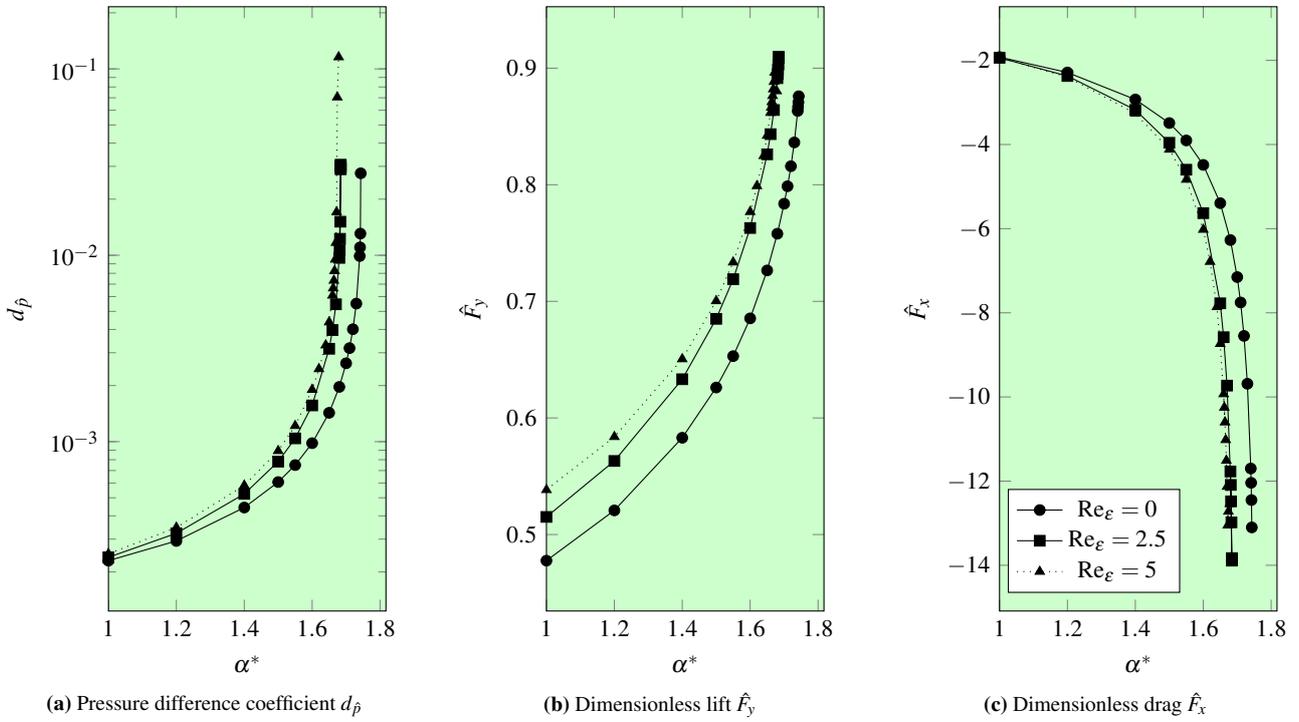
\begin{figure*}
\centering
\subfloat[Pressure difference coefficient $d_{\hat{p}}$]{%
\centering
	\begin{tikzpicture}
		\begin{semilogyaxis}[
			width={0.3\textwidth}, height={0.4\textheight},
			xlabel=$\alpha^*$,  
			ylabel=$d_{\hat{p}}$,
			unbounded coords=discard,
			xmin=0.0,
			font=\small,
			axis background/.style={fill=green!20},	
			legend pos=north west,
			cycle list name=mark list* 
		]
		\addplot+ table[x index=3,y index=5, smooth, solid, mark=.] {results/1602a-dp-eps-alpha/dp-e5e-3-G0.0-Re0.0.txt};
		\addplot+ table[x index=3,y index=5, smooth, solid, mark=.] {results/1602a-dp-eps-alpha/dp-e5e-3-G0.2-Re0.0.txt};
		\addplot+ table[x index=3,y index=5, smooth, solid, mark=.] {results/1602a-dp-eps-alpha/dp-e5e-3-G0.5-Re0.0.txt};
		\end{semilogyaxis}
	\end{tikzpicture}
}\hspace{0.04\textwidth}\subfloat[Dimensionless lift $\hat{F}_y$]{%
	\centering
	\begin{tikzpicture}
		\begin{axis}[
			width={0.3\textwidth}, height={0.4\textheight},
			xlabel=$\alpha^*$,  
			ylabel=$\hat{F}_y$,
			unbounded coords=discard,
			xmin=0.0,
			ymin=0,
			font=\small,
			axis background/.style={fill=green!20},	
			legend style={anchor=south west,at={(0.1,0.05)},reverse legend}, 
			cycle list name=mark list*, 
		]
		\addplot+ table[x index=3,y index=7, smooth, solid, mark=.] {results/1602a-dp-eps-alpha/dp-e5e-3-G0.0-Re0.0.txt};
			\addlegendentry{ $G^*=0$ };
		\addplot+ table[x index=3,y index=7, smooth, solid, mark=.] {results/1602a-dp-eps-alpha/dp-e5e-3-G0.2-Re0.0.txt};
			\addlegendentry{ $G^*=0.2$ };
		\addplot+ table[x index=3,y index=7, smooth, solid, mark=.] {results/1602a-dp-eps-alpha/dp-e5e-3-G0.5-Re0.0.txt};
			\addlegendentry{ $G^*=0.5$ };
		\end{axis}
	\end{tikzpicture}
}\hspace{0.04\textwidth}\subfloat[Dimensionless drag $\hat{F}_x$]{
	\centering
	\begin{tikzpicture}
		\begin{axis}[
			width={0.3\textwidth}, height={0.4\textheight},
			xlabel=$\alpha^*$,  
			ylabel=$\hat{F}_x$,
			unbounded coords=discard,
			xmin=0.0,
			font=\small,
			axis background/.style={fill=green!20},	
			legend pos=south west,
			cycle list name=mark list* 
		]
		\addplot+ table[x index=3,y index=8, smooth, solid, mark=.] {results/1602a-dp-eps-alpha/dp-e5e-3-G0.0-Re0.0.txt};
			\addlegendentry{ $G^*=0$ };
		\addplot+ table[x index=3,y index=8, smooth, solid, mark=.] {results/1602a-dp-eps-alpha/dp-e5e-3-G0.2-Re0.0.txt};
			\addlegendentry{ $G^*=0.2$ };
		\addplot+ table[x index=3,y index=8, smooth, solid, mark=.] {results/1602a-dp-eps-alpha/dp-e5e-3-G0.5-Re0.0.txt};
			\addlegendentry{ $G^*=0.5$ };
		\legend{}
		\end{axis}
	\end{tikzpicture}
}
\caption{Dimensionless force $\hat{\vect{F}}$ and pressure difference coeficient $d_{\hat{p}}$, 
	variation with $G^*$ and $\alpha^*$
	($h_2/h_1=2$, $\Rey_\eps=0$, $\eps=0.005$ and $r=3/2$, $\beta/\alpha=2$)} 
\label{plot:dp-eps-alpha-G}
\end{figure*}
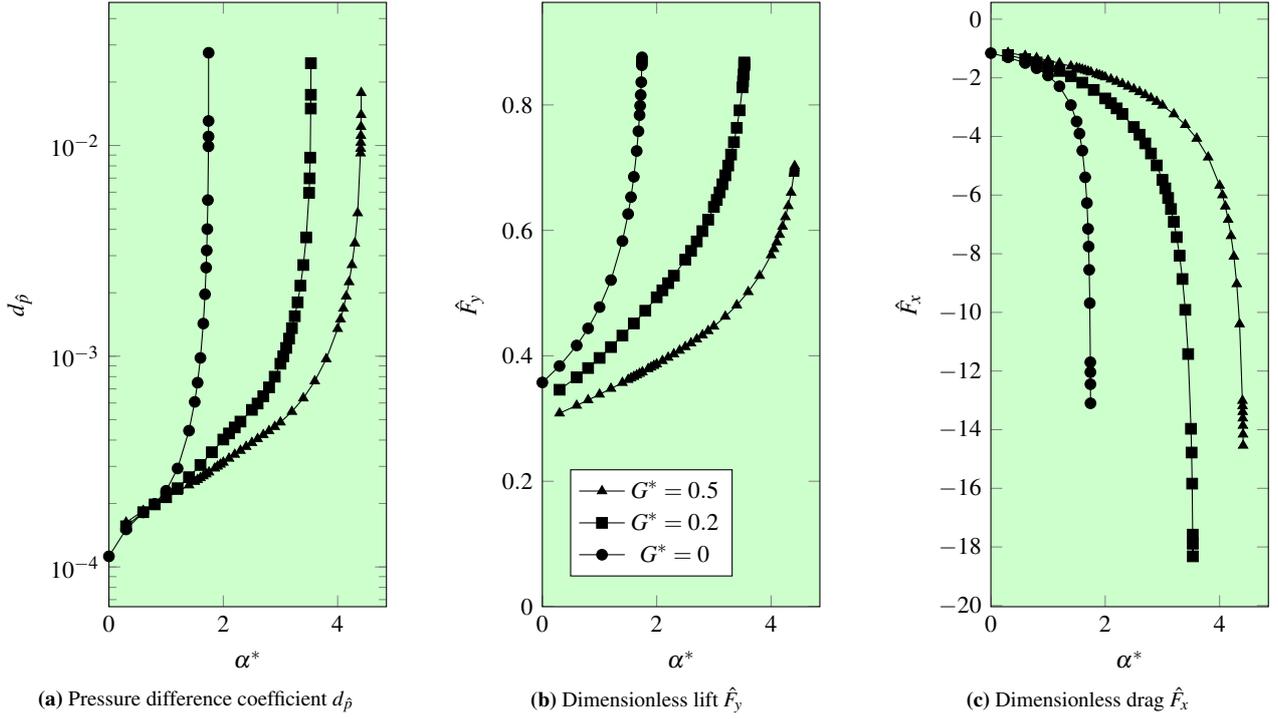
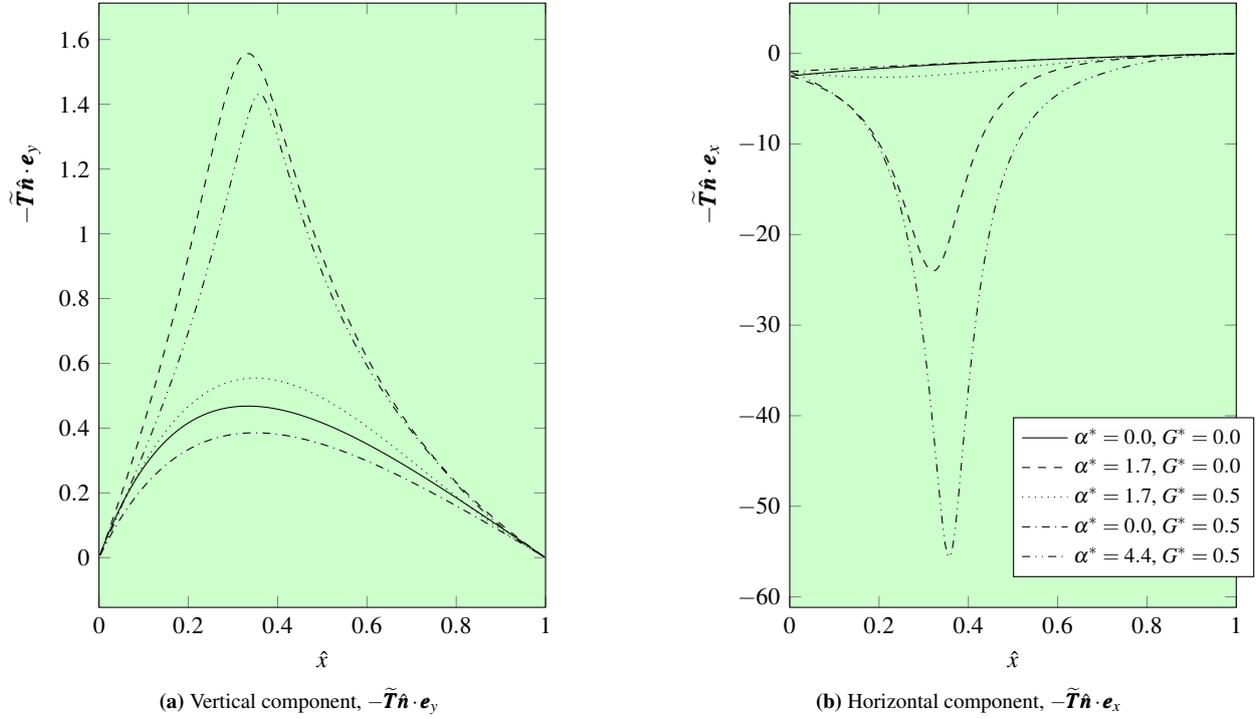
\begin{figure*}
\centering
\subfloat[Vertical component, $-\widetilde{\tens{T}}\hat{\vect{n}}\cdot\vect{e}_y$]{%
\begin{minipage}{0.45\textwidth}
	\begin{tikzpicture}
	\vspace{1em}%
		\begin{axis}[
			xlabel={ $\hat{x}$ },  
			ylabel={ $-\widetilde{\tens{T}}\hat{\vect{n}}\cdot\vect{e}_y$ },
			width={ 0.95\textwidth }, height={ 0.4\textheight },
			every axis y label/.style=
				{at={(ticklabel cs:0.7)},rotate=90,anchor=near ticklabel},
			unbounded coords=discard,
			xmin=0.0, xmax=1.0,
			font=\small,
			axis background/.style={fill=green!20},	
			legend pos=outer north east,
			cycle list name=linestyles*, 
		]
		\addplot+ table[x index=0,y index=3, smooth, solid, mark=none,col sep=space,trim cells=true] {results/1602a-dp-eps-alpha/00001m6.Tnxy-slider.0001};
		  \addlegendentry{ $\alpha^*=0.0$, $G^*=0.0$ }
		\addplot+ table[x index=0,y index=3, smooth, solid, mark=none,col sep=space,trim cells=true] {results/1602a-dp-eps-alpha/00013m6.Tnxy-slider.0001};
		  \addlegendentry{ $\alpha^*=1.7$, $G^*=0.0$ }
		\addplot+ table[x index=0,y index=3, smooth, solid, mark=none,col sep=space,trim cells=true] {results/1602a-dp-eps-alpha/18012m6.Tnxy-slider.0001};
		  \addlegendentry{ $\alpha^*=1.7$, $G^*=0.5$ }
		\addplot+ table[x index=0,y index=3, smooth, solid, mark=none,col sep=space,trim cells=true] {results/1602a-dp-eps-alpha/30000m6.Tnxy-slider.0001};
		  \addlegendentry{ $\alpha^*=0.0$, $G^*=0.5$ }
		\addplot+ table[x index=0,y index=3, smooth, solid, mark=none,col sep=space,trim cells=true] {results/1602a-dp-eps-alpha/11036m6.Tnxy-slider.0001};
		  \addlegendentry{ $\alpha^*=4.4$, $G^*=0.5$ }
		\legend{}
		\end{axis}
	\end{tikzpicture}
\end{minipage}}
\hspace{0.05\textwidth}
\subfloat[Horizontal component, $-\widetilde{\tens{T}}\hat{\vect{n}}\cdot\vect{e}_x$]{%
\begin{minipage}{0.45\textwidth}
	\centering
	\begin{tikzpicture}
	\vspace{1em}%
		\begin{axis}[
			xlabel={ $\hat{x}$ },  
			ylabel={ $-\widetilde{\tens{T}}\hat{\vect{n}}\cdot\vect{e}_x$ },
			width={ 0.95\textwidth }, height={ 0.4\textheight },
			every axis y label/.style=
				{at={(ticklabel cs:0.7)},rotate=90,anchor=near ticklabel},
			unbounded coords=discard,
			xmin=0.0, xmax=1.0,
			font=\small,
			axis background/.style={fill=green!20},	
			legend style={anchor=south west,at={(0.5,0.05)},font=\footnotesize}, 
			cycle list name=linestyles*, 
		]
		\addplot+ table[x index=0,y index=2, smooth, solid, mark=none,col sep=space,trim cells=true] {results/1602a-dp-eps-alpha/00001m6.Tnxy-slider.0001};
		  \addlegendentry{ $\alpha^*=0.0$, $G^*=0.0$ }
		\addplot+ table[x index=0,y index=2, smooth, solid, mark=none,col sep=space,trim cells=true] {results/1602a-dp-eps-alpha/00013m6.Tnxy-slider.0001};
		  \addlegendentry{ $\alpha^*=1.7$, $G^*=0.0$ }
		\addplot+ table[x index=0,y index=2, smooth, solid, mark=none,col sep=space,trim cells=true] {results/1602a-dp-eps-alpha/18012m6.Tnxy-slider.0001};
		  \addlegendentry{ $\alpha^*=1.7$, $G^*=0.5$ }
		\addplot+ table[x index=0,y index=2, smooth, solid, mark=none,col sep=space,trim cells=true] {results/1602a-dp-eps-alpha/30000m6.Tnxy-slider.0001};
		  \addlegendentry{ $\alpha^*=0.0$, $G^*=0.5$ }
		\addplot+ table[x index=0,y index=2, smooth, solid, mark=none,col sep=space,trim cells=true] {results/1602a-dp-eps-alpha/11036m6.Tnxy-slider.0001};
		  \addlegendentry{ $\alpha^*=4.4$, $G^*=0.5$ }
		\end{axis}
	\end{tikzpicture}
\end{minipage}}
\caption{Traction along the slider surface $\hat{\Gamma}_{\mathrm{slider}}$
	for selected $\alpha^*$ and $G^*$
	($h_2/h_1=2$, $\Rey_\eps=0$, $\eps=0.005$, and $r=3/2$, $\beta/\alpha=2$)}
\label{plot:comparison-A174}
\end{figure*}
\begin{figure*}
\centering
\subfloat[$\alpha^*=1.7$, $G^*=0.0$]{
	\includegraphics[width=0.45\textwidth,trim=200 300 50 250,clip=true]{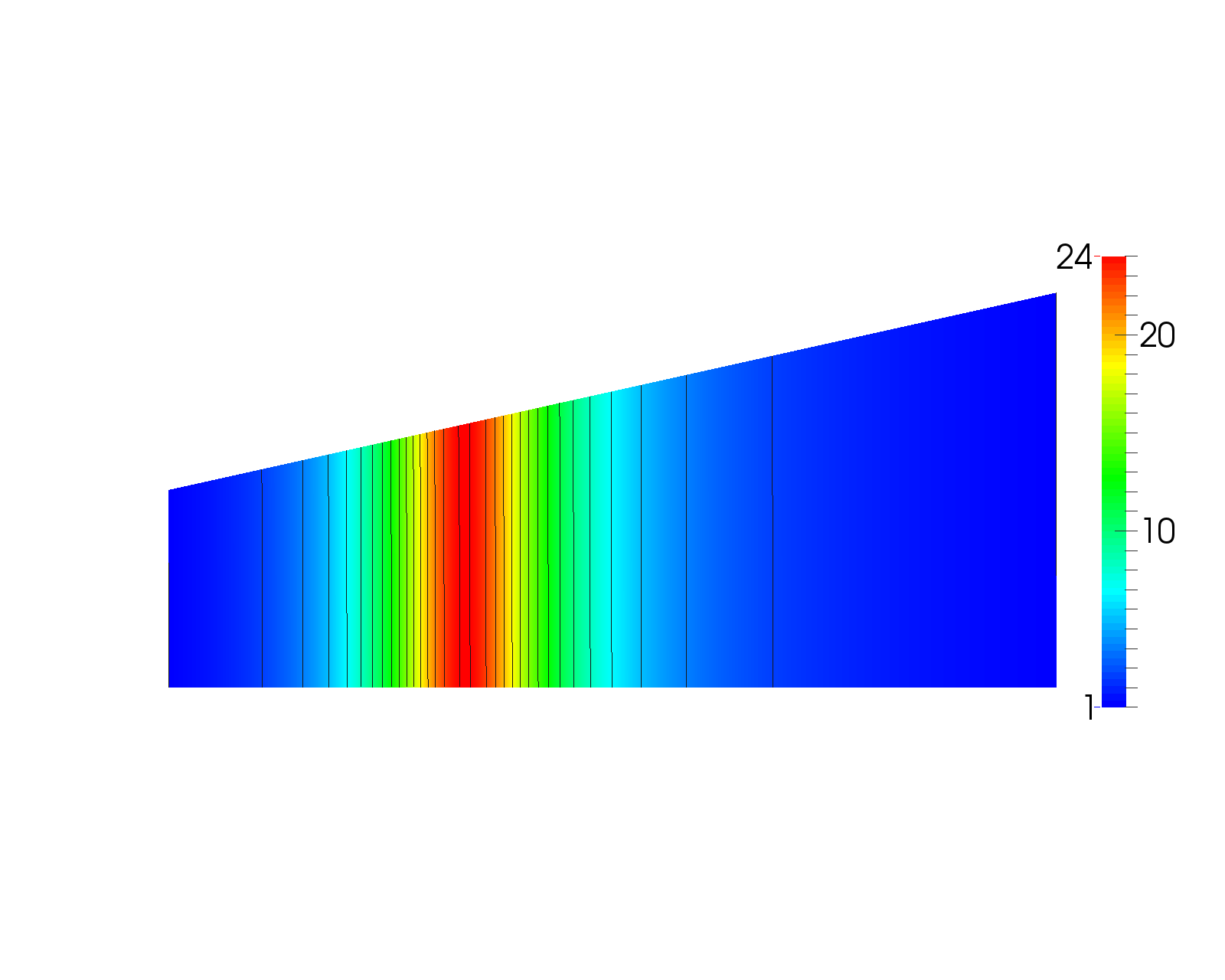}
}
\hspace{0.05\textwidth}
\subfloat[$\alpha^*=1.7$, $G^*=0.5$]{
	\includegraphics[width=0.45\textwidth,trim=200 300 50 250,clip=true]{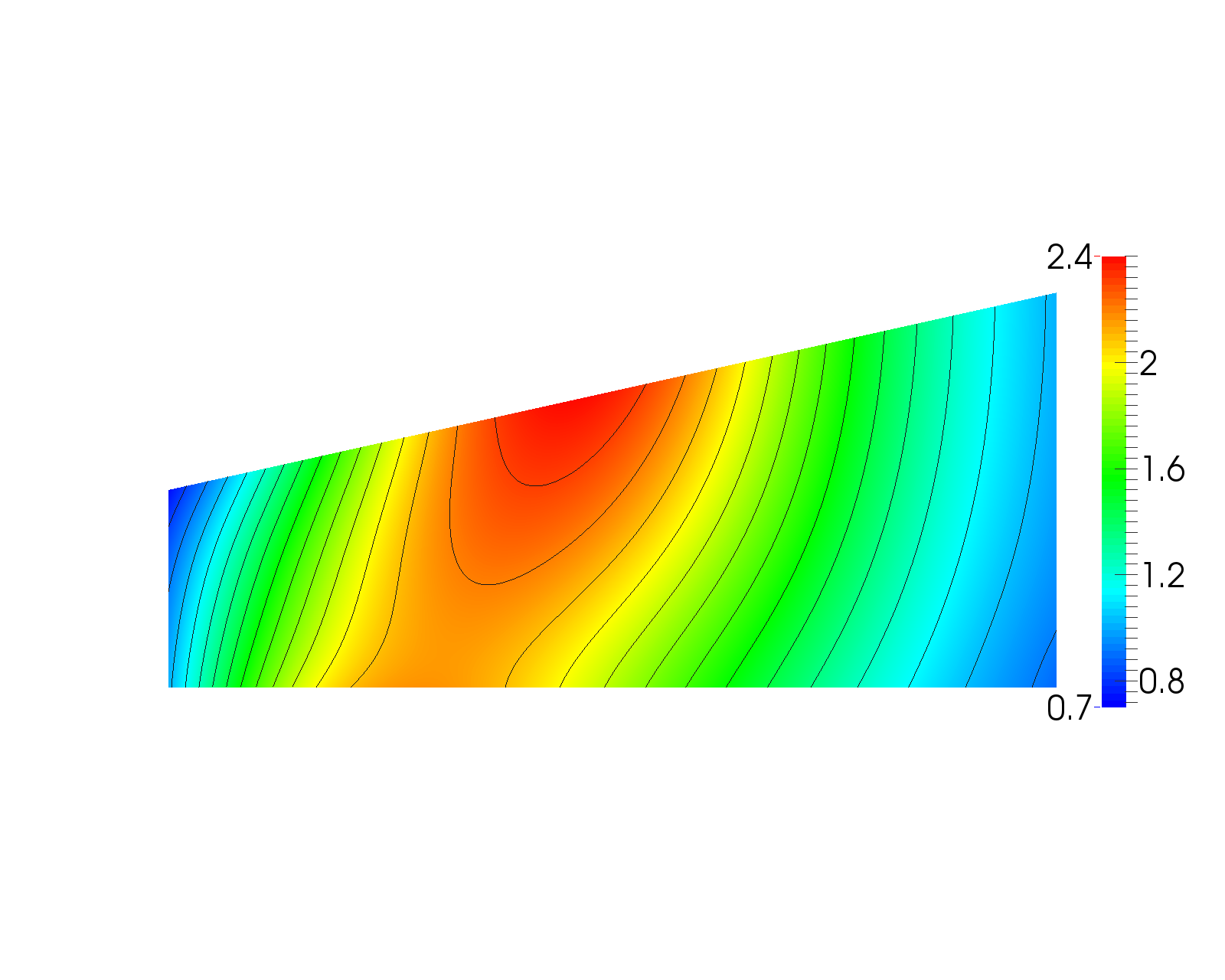}
}
\\ 
\subfloat[$\alpha^*=0.0$, $G^*=0.5$]{
	\includegraphics[width=0.45\textwidth,trim=200 300 50 250,clip=true]{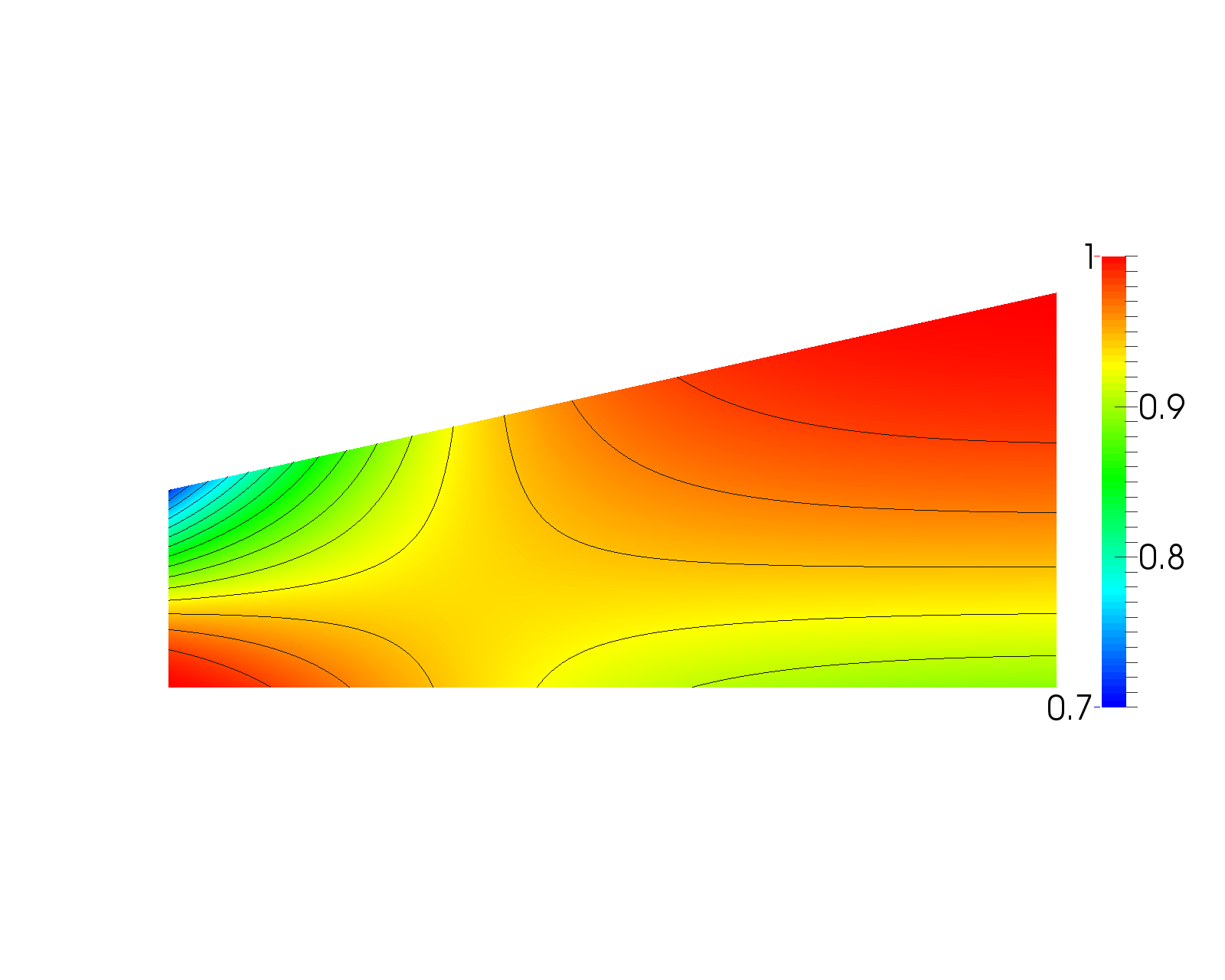}
}
\hspace{0.05\textwidth}
\subfloat[$\alpha^*=4.4$, $G^*=0.5$]{
	\includegraphics[width=0.45\textwidth,trim=200 300 50 250,clip=true]{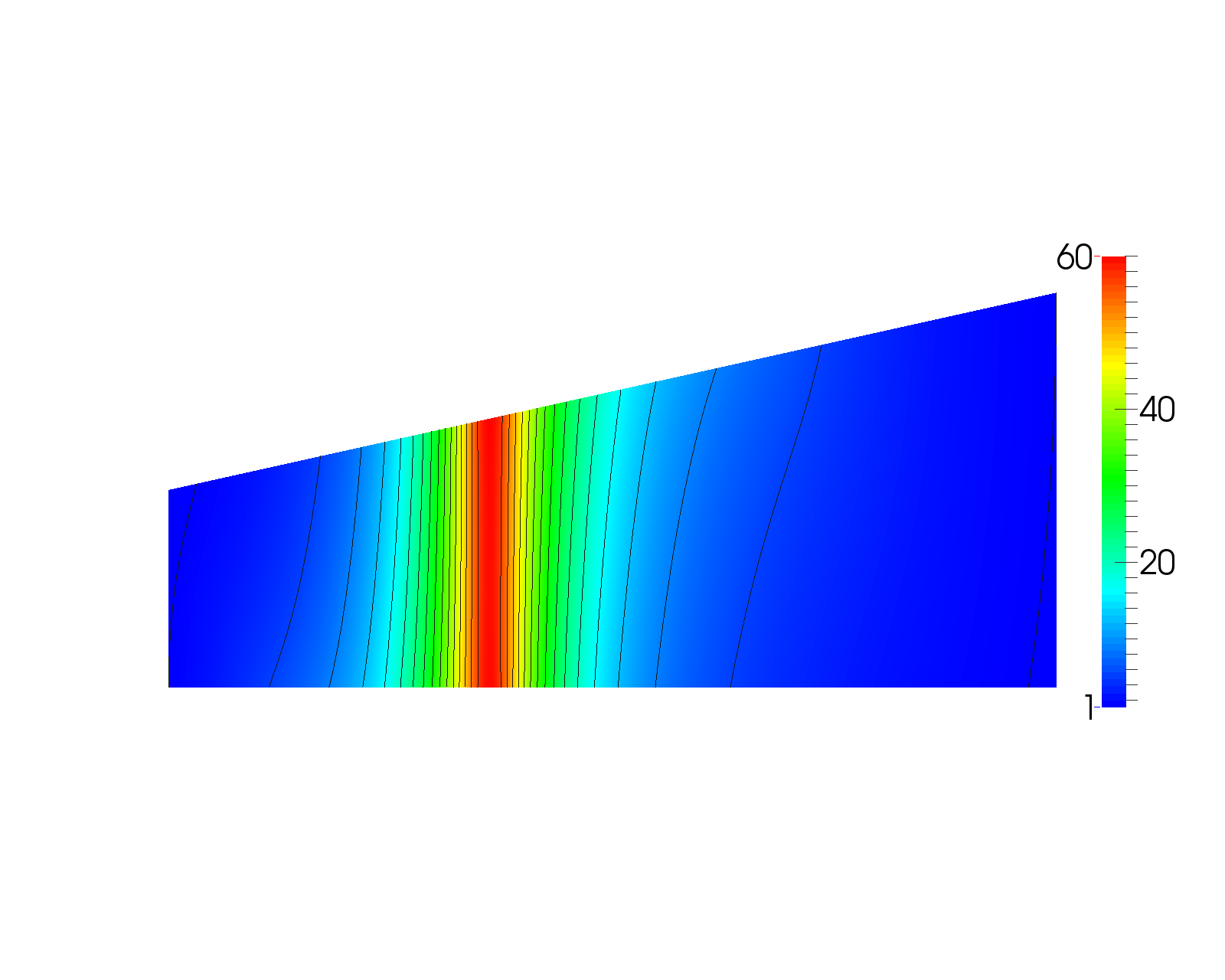}
}
\caption{ Dimensionless viscosity $\hat\eta$ 
	for selected $\alpha^*$ and $G^*$
	($h_2/h_1$, $\Rey_\eps=0$, $\eps=0.005$, and $r=3/2$, $\beta/\alpha=2$)}
\label{fig:comparison-visco}
\end{figure*}
Within a~unidirectional flow, such as Couette flow or plane Poiseuille flow, 
of a~Navier--Stokes fluid or a~fluid with shear rate dependent viscosity,
in the absence of body forces, the pressure gradient is either trivial or its direction
is that of the flow.
In the context of lubrication flows, the almost unidirectional flow within the thin film then corresponds
to negligible pressure variations across the film.
The situation differs significantly if the viscosity varies with the pressure.
This was well documented for the Couette and Poiseuille plane flows, 
see e.g.~\cite{BairKhonsariWiner_1998,HronMalekRaj_2001,HronMalekPrusaRaj_2009}.
In fact, for the exponential pressure--viscosity relation $\eta=e^{\alpha p}$
no such unidirectional flow can be found
(except, interestingly enough, the case with a~cross-flow pressure gradient due to the gravitational force,
see~\cite{MalekRaj_2007}).
It was pointed out in~\cite{RajSzeri_2003} that the cross-flow pressure gradient induced within 
the lubrication flow in the piezoviscous regime gives rise to an~additional term in the 
Reynolds approximation equation, see also~\cite{BayadaEtal_2013,GustafssonRajEtal_2015}.

The results of numerical computations presented in \Figurep{dp-eps-alpha}
reveal how the pressure differences appear with increasing $\alpha^*$, for different values of~$\eps$.
Notice again that each plotted curve ends at certain critical value of $\alpha^*$, for which (and all the higher values)
the condition~\eqref{eq:dSdp} is violated and the numerical scheme fails. 
An increase in the coefficient $d_{\bar{p}}$ by as much as two orders of magnitude,
when compared to the Navier--Stokes fluid at given $\eps$, 
can be observed before such critical $\alpha^*$ is reached.
Simultaneously, a~rapid increase of the maximal dimensionless pressure (not visualised) and
both components of the resulting force $\hat{\vect{F}}$ appear. 
Note how the critical values of $\alpha^*$ differ with $\eps$, say for $\eps>0.01$,
as can be read from \Figurep{dp-eps-alpha}.
We observe that for $\alpha^*>0$ the resulting dimensionless force is 
more sensitive to $\eps$ than it was shown for a~Navier--Stokes lubricant, cf.~\Figurep{draglift}.

\subsection{The shear-thinning and inertial effects}
We complete the presentation of the numerical computations by including a~sample of results
with shear-thinning, i.e.\ with $G^*>0$, and the results for $\Rey_\eps>0$, 
in addition to pressure-thickening.
The observed coefficient $d_{\bar{p}}$ and the resulting dimensionless force 
are again plotted in Figs.~\ref{plot:dp-eps-alpha-Rey} and~\ref{plot:dp-eps-alpha-G}.
For the simplicity of presentation we keep $r=3/2$ and $\beta/\alpha=2$ and 
only present the results for $\eps=0.005$.

For $\alpha^*>0$ and small values of~$\eps$, the numerical simulations for positive $\Rey_\eps$
are more demanding in comparison to the case $\alpha^*=0$.
In contrast to the results presented in \Figurep{Dp--a}, we observed that
the discrete solutions for, let us say, $\alpha^*>1.5$ with $\Rey_\eps=5$ or higher 
remain mesh--dependent for regular mesh refinements as fine as $\hat{h}\sim2^{-6}$
(corresponding to $136\,194$ degrees of freedom).
The comparison of the resulting $d_{\bar{p}}$ and $\hat{\vect{F}}$ for $\Rey_\eps=0$ and $2.5$
is plotted in \Figurep{dp-eps-alpha-Rey}, illustrating how 
the increased modified Reynolds number leads (by means of increasing the generated pressure peak)
to the increased dimensionless force. 
The approximation obtained for $\Rey_\eps=5$ is included as the dotted line.
Further study of the \emph{combined} effects of pressure-thickening and higher Reynolds numbers 
would require some additional care which we exclude from the current presentation.

With the shear-thinning taking effect, the growth of the maximal pressure and viscosity
with increasing $\alpha^*$ is postponed, thus increasing significantly the observed critical 
value of $\alpha^*$ for which \eqref{eq:dSdp} is violated within the resulting flow.
More detailed comparison is provided in \Figurep{comparison-A174}, 
where the distribution of the dimensionless traction along the slider surface is plotted
for five combinations of $\alpha^*$ and $G^*$.
All these results are for $\Rey_\eps=0$ and $\eps=0.005$.

For reference, the solid line is plotted in \Figurep{comparison-A174}
representing a~constant viscosity lubricant.
The dashed line then shows the pure piezoviscous regime with $\alpha^*=1.7$,
displaying the large sharp pressure peak on the left plot and the increased friction contributions
due to the corresponding peak in the viscosity, on the right-hand side plot.
With the same $\alpha^*$ but with $G^*=0.5$, as can be read from the dotted line, 
the effect of piezoviscous response is largely counteracted by shear-thinning.
For comparison, the case of $G^*=0.5$ but $\alpha^*=0$ is also included, 
showing much lesser variation due to shear-thinning in the case of $\alpha^*=0$, 
when compared to the piezoviscous regime for $\alpha^*=1.7$.

Finally, we include the dashed-double-dotted plot for the case $\alpha^*=4.4$ and $G^*=0.5$,
to emphasize the difference in influence of these two parameters on the two components of 
the resulting force:
Note that for $\alpha^*=4.4$, $G^*=0.5$ the vertical traction (and so the pressure peak)
almost reaches the values for the pure piezoviscous $\alpha^*=1.7$, $G^*=0.0$, 
the peak being slightly sharper and shifted towards the inlet.
By contrast, significantly larger horizontal traction is observed.

The distribution of the dimensionless viscosity in $\hat\Omega$ 
is presented by means of contour plots in \Figure{comparison-visco},
for the same four cases: 
(a) the pure piezoviscous case $\alpha^*=1.7$, $G^*=0$, showing a~sharp viscosity peak
reaching the maximum $\hat{\eta}=24$,
(b) the case $\alpha^*=1.7$, $G^*=0.5$, where the viscosity peak is an~order of magnitude lower
(which is also accompanied by the significantly lower pressure peak),
(c) the pure shear-thinning case $\alpha^*=0$, $G^*=0.5$, and finally
(d) the case $\alpha^*=4.4$, $G^*=0.5$, 
showing the viscosity peak reaching the maximum $\hat\eta=60$ 
as well as the variation of the viscosity due to the velocity gradient in the remaining parts of the domain.

\section{Conclusion}
Based on the numerical computations that have been carried out, 
we conclude that the finite element solution
for the planar steady isothermal flow of an~incompressible fluid with 
pressure and shear rate dependent viscosity  can be obtained as long
as the condition \eqref{eq:dSdp} is satisfied.
Note that the condition~\eqref{eq:dSdp} supplemented by certain additional assumptions
also guarantees the existence of solutions to the full equations governing the flows of the fluids under consideration.
Once the condition is violated, i.e.~if the pressure \emph{or} shear rate
reach values larger than some critical value, we were unable to obtain any numerical solution.

As the parameters approach the critical case, the rapid growth of 
the quantities tracked in the plane slider simulations,
such as the maximal values of the pressure and viscosity 
and the force acted on the solid surfaces, were observed.

In particular, we have documented the implications of cutting the viscosity off 
above a~given threshold of pressure: the technique does not guarantee
convergence and, once the cut-off takes effect, the results depend critically 
on the artificial threshold parameter.
The effect is particularly pronounced when the overall friction 
(i.e.~the tangential part of the traction observed on the solid walls) is considered.

In the range of parameters where the unaltered viscosity can be considered,
we discussed the resulting plane slider flow for a~number of combinations 
of the dimensionless parameters related to the pressure-thickening, shear-thinning,
inertia and geometry. 
In particular, we tracked the force acting on the slider surface as it
varies with the dimensionless pressure--viscosity coefficient $\alpha^*$
for different parameters $\eps$, where $\eps\searrow0$ would represent the
lubrication approximation limit, and with different parameters $G^*$ related
to the activation of the shear-thinning response.

In order to study the variations of pressure and other quan\-tities accross the film,
the boundary conditions taken on the artificial (inflow and outflow) boundaries 
needed to be discussed.
We have observed that the condition \eqref{eq:bc-artificial},
derived in \Section{bc} based on the \emph{do-nothing} condition used for Navier--Stokes fluid, 
is appropriate for the problem under consideration.
In contrast to, e.g., constant traction being prescribed, 
we observed smooth solutions without any artifacts in the pressure or viscosity field
in the vicinity of the artificial boundaries.

We have displayed how the pressure variations across the film 
appear within the flow due to pressure-thickening.
The results may imply that the lubrication assumptions are violated 
by the piezoviscous lubricant.
This assertion has been made already by researchers working with the 
Reynolds approximation, and it was our hope 
to provide a~numerical validation to the recently derived corrections of 
Reynolds equation. 
Unfortunately, as the appearance of pressure variations is in conjunction
with the change of the structure in the momentum equation,
the most important comparison would require one 
to find a~numerical solution to the problem in the case, where the condition \eqref{eq:dSdp} is violated.
This represents a~challenging open problem in computational fluid dynamics of incompressible fluids.
To the best of our knowledge, no numerical solutions have been reported in the literature so far
that would reach beyond \eqref{eq:dSdp}.
Similarly, there are no theoretical results either, concerning the existence of such a~solution.


\bibliographystyle{spbasic}      


\end{document}